\documentclass[twocolumn,secnumarabic,amssymb,nobibnotes,aps,prd,superscriptaddress,nofootinbib]{revtex4-1}

\pdfoutput=1

\usepackage{epsfig}
\usepackage{graphics}
\usepackage{graphicx}
\usepackage{amsmath,amsfonts,amssymb,amsthm}
\usepackage[usenames,dvipsnames]{xcolor}
\usepackage{wasysym}
\usepackage{times}
\usepackage{gensymb}
\usepackage{appendix}
\usepackage{listings}
\usepackage{url}
\usepackage[normalem]{ulem}
\usepackage{alltt}
\usepackage[colorlinks]{hyperref}
\usepackage{cleveref}
\usepackage{longtable}
\usepackage{enumitem}
\usepackage{orcidlink}
\setlist{nosep}
\usepackage{color}
\usepackage{calc}
\usepackage{tensor}
\usepackage{bm}
\usepackage{multirow}
\usepackage[varg]{txfonts}
\usepackage{float}
\usepackage{dcolumn}
\usepackage[nolist,nohyperlinks]{acronym}
\usepackage{xspace}
\usepackage[english]{babel}
\usepackage[abs]{overpic}
\usepackage{pict2e}
\usepackage[caption=false]{subfig} 
\allowdisplaybreaks[1]
\usepackage[utf8]{inputenc}
\usepackage{gensymb}
\usepackage{bm}
\usepackage{stackengine}
\usepackage{boldline,multirow}
\usepackage{braket}
\usepackage{longtable}
\usepackage{makecell}
\usepackage{stackengine}

\usepackage{tabularx}
\usepackage{rotating}
\usepackage{booktabs}

\newcommand{\AEI}{Max Planck Institute for Gravitational Physics (Albert Einstein Institute), Am M\"uhlenberg 1, Potsdam 14476, Germany}
\newcommand{\Maryland}{Department of Physics, University of Maryland, College Park, MD 20742, USA}
\newcommand{\URI}{Department of Physics, East Hall, University of Rhode Island, Kingston, RI 02881, USA}
\newcommand{\URICCR}{Center for Computational Research, Tyler Hall, University of Rhode Island, Kingston, RI 02881, USA}

\definecolor{citecolor}{HTML}{00707e}
\definecolor{linkcolor}{HTML}{f94730}
\definecolor{urlcolor}{HTML}{1c4484}

\hypersetup{citecolor=citecolor, linkcolor=linkcolor, urlcolor=urlcolor}

\newcommand{\seob}{\texttt{SEOBNRv5HM}}

\begin{document}

\title{Accounting for numerical-relativity calibration uncertainty \\
	in gravitational-wave modeling and inference}

\author{Lorenzo Pompili\,\orcidlink{0000-0002-0710-6778}}
\email{lorenzo.pompili@aei.mpg.de}
\affiliation{\AEI}

\author{Alessandra Buonanno\,\orcidlink{0000-0002-5433-1409}}
\affiliation{\AEI}
\affiliation{\Maryland}

\author{Michael Pürrer\,\orcidlink{0000-0002-3329-9788}}
\affiliation{\URI}
\affiliation{\URICCR}
\affiliation{\AEI}

\date{\today}

\begin{abstract}
	The increasing sensitivity of current and upcoming
	gravitational-wave (GW) detectors
	poses stringent requirements on the accuracy of the GW models used for data analysis.
	If these requirements are not met, systematic errors could dominate over
	statistical uncertainties, hindering our ability to extract
	astrophysical and cosmological information, and conduct precise tests of general relativity. In this work, we present a novel method to mitigate the impact of waveform-systematic errors, by incorporating and marginalizing over waveform-uncertainty estimates, which are modeled as probability distributions for the numerical-relativity
	calibration parameters of effective-one-body waveform models.
	By analyzing simulated GW signals of loud ``golden'' binary-black-hole systems, we show that our method significantly reduces biases in the recovered parameters, highlighting its potential to improve the robustness of GW parameter estimation with upcoming observing runs and next-generation ground-based facilities, such as the Einstein Telescope and Cosmic Explorer.
\end{abstract}

\maketitle


\section{Introduction} Gravitational-wave (GW) astronomy has rapidly
advanced since the first GW observation from a binary-black-hole
(BBH) merger~\cite{LIGOScientific:2016aoc}, with about 100 compact-binary signals
detected by the LIGO-Virgo-KAGRA (LVK)
collaboration~\cite{LIGOScientific:2014pky, VIRGO:2014yos,
	KAGRA:2020tym, KAGRA:2021vkt} and by independent
analyses~\cite{Nitz:2021zwj, Mehta:2023zlk}. These observations have
shed light on the mass and spin distributions of BHs
and neutron stars (NSs)~\cite{KAGRA:2021duu}, improved constraints on
the nuclear equation of state of NSs~\cite{LIGOScientific:2018cki},
obtained independent measurements of cosmological
parameters~\cite{LIGOScientific:2017adf, LIGOScientific:2021aug}, and
enabled tests of general relativity (GR) in the highly dynamical
and strong-field regime~\cite{LIGOScientific:2016lio, LIGOScientific:2021sio}.

Advancements in gravitational waveform modeling (e.g., see~\cite{Buonanno:2006ui, Damour:2007yf, Ajith:2007qp,
	Taracchini:2012ig, Khan:2015jqa, Bohe:2016gbl, Blackman:2017pcm,
	Nagar:2018zoe, Varma:2019csw, Ossokine:2020kjp, Pratten:2020ceb,
	Estelles:2021gvs, Gamba:2021ydi, Nagar:2023zxh, Pompili:2023tna,
	Ramos-Buades:2023ehm, Thompson:2023ase,
	LISAConsortiumWaveformWorkingGroup:2023arg}) have been crucial to
the success of this endeavor, with waveform models being sufficiently
accurate to analyze most detected
signals~\cite{LIGOScientific:2016ebw, Purrer:2019jcp}, with only a few exceptions where different models recovered noticeably different parameters~\cite{KAGRA:2021vkt, Hu:2022rjq, Islam:2023zzj}.
Next-generation (XG) GW detectors on the ground, such as the Einstein
Telescope (ET)~\cite{Punturo:2010zz} and Cosmic Explorer
(CE)~\cite{Reitze:2019iox}, and the space-based
LISA~\cite{LISA:2017pwj}, will offer unprecedented sensitivity, with
signal-to-noise ratios (SNR) up to two orders of magnitude higher than
current detectors~\cite{Borhanian:2022czq}.  This increased
sensitivity makes waveform accuracy increasingly critical, as statistical uncertainties approach the systematic
biases of the GW approximant models. Several studies predict
severe biases in parameter estimation (PE) due to mismodeling errors with the
upcoming fifth LVK observational run (O5) and
XG detectors~\cite{Purrer:2019jcp, Gamba:2020wgg, Kunert:2021hgm,
	Owen:2023mid, Ramos-Buades:2023ehm, Dhani:2024jja, Kapil:2024zdn}, and the risk of false
GR deviations~\cite{Hu:2022bji, Toubiana:2023cwr, Gupta:2024gun,
	Foo:2024exr}.  Even numerical-relativity (NR)
simulations~\cite{Pretorius:2005gq, Campanelli:2005dd, Baker:2005vv},
the benchmark for generating highly accurate waveforms, are expected
to fall short of the required accuracy~\cite{Ferguson:2020xnm,
	Jan:2023raq}.

Waveform models for compact binaries combine
analytical and NR results. The effective-one-body (EOB)
approach~\cite{Buonanno:1998gg, Buonanno:2000ef, Damour:2000we,
	Damour:2001tu, Buonanno:2005xu} synthesizes and resums analytical
information from various perturbative methods for solving Einstein's
equations, with current EOB waveform models primarily relying on
resummations of the post-Newtonian (PN)~\cite{Futamase:2007zz,
	Blanchet:2013haa} expansion, applicable in the weak-field and
small-velocity limit. To improve agreement with NR simulations,
higher-order PN coefficients, which are currently unknown, are often
calibrated using NR data (e.g., see~\cite{Buonanno:2007pf, Damour:2007yf,
	Bohe:2016gbl, Nagar:2018zoe, Nagar:2023zxh, Pompili:2023tna}).

A possible way to improve waveform accuracy is to
include higher-order analytical information, while pushing calibration
parameters to even higher orders. However, this requires careful
studies on how to incorporate and resum this
information (e.g., see~\cite{Nagar:2021xnh, Khalil:2023kep,Pompili:2023tna, Nagar:2023zxh,
	Nagar:2024dzj}), and improvements in the strong-field regime are not
always guaranteed. Leveraging novel perturbative methods, such as the gravitational
self-force (GSF)~\cite{Damour:2009sm,Akcay:2012ea, LeTiec:2011dp,Antonelli:2019fmq,
	vandeMeent:2023ols} and the post-Minkowskian
(PM)~\cite{Damour:2016gwp, Antonelli:2019ytb, Khalil:2022ylj,
	Damour:2022ybd, Rettegno:2023ghr, Buonanno:2024vkx,
	Buonanno:2024byg} approximations, offers another promising
avenue. Finally, improving the NR calibration of waveform models by
employing longer, higher-resolution simulations covering more of the
parameter space is crucial~\cite{Pan:2013tva,Bohe:2016gbl}.

A common approach to address waveform systematics is to combine predictions from different waveform models.
This can be done either by mixing posterior samples from various approximants~
\cite{LIGOScientific:2016ebw, KAGRA:2021vkt, Ashton:2019leq, Jan:2020bdz},
or by sampling over a set of models treated as hyperparameters in the analysis~
\cite{Ashton:2021cub, Hoy:2022tst, Puecher:2023rxw, Hoy:2024vpc}.
However, if none of the models employed is accurate enough,
these methods are not guaranteed to avoid biases in the inferred parameter estimates.
A complementary approach to avoid biased inference involves
integrating error estimates directly into waveform models. Proposed
methods include using Gaussian process regression (GPR) to interpolate
either waveform residuals or directly NR
waveforms~\cite{Moore:2014pda, Gair:2015nga, Moore:2015sza,
	Doctor:2017csx, Williams:2019vub, Andrade:2023sal, Khan:2024whs},
using frequency-dependent amplitude and phase corrections, as in the
case of detector-calibration uncertainty~\cite{Edelman:2020aqj, Read:2023hkv}, or
introducing additional higher-order parameters designed to capture
currently unknown PN terms~\cite{Owen:2023mid}.
Similar methods have also been applied to model the
postmerger phase of NS coalescences~\cite{Breschi:2021xrx, Breschi:2022xnc},
as well as in kilonova modeling and inference~\cite{Breschi:2024qlc}.
In these approaches, waveform modeling uncertainties are accounted for by marginalizing
over these additional degrees of freedom during the inference
process. While this may lead to less precise parameter estimates
(e.g., wider posterior distributions), it should improve
their robustness, ensuring they remain reliable even in the presence
of systematic modeling errors.

Nonetheless, most studies attempting to incorporate error estimates
into waveform models often
involved simplified physical descriptions (e.g., nonspinning
binaries~\cite{Owen:2023mid, Khan:2024whs}), and did not assess the
impact of this approach on the analysis of realistic signals that
could be observed with upcoming runs and XG detectors. In this study, we
address this gap. As proof of principle, we employ
a state-of-the-art quasicircular aligned-spin multipolar EOB model \seob~\cite{Pompili:2023tna,
	Khalil:2023kep, vandeMeent:2023ols, Mihaylov:2023bkc}, and analyze
synthetic signals from NR surrogate waveforms, hybridized to PN-EOB ones
(\texttt{NRHybSur3dq8}~\cite{Varma:2018mmi}), for loud ``golden'' BBH
systems at SNRs consistent with current and future observations, using
full Bayesian inference.

Similarly to Refs.~\cite{Doctor:2017csx, Khan:2024whs},
we propose a semiparametric probabilistic model for
the GW signal from a BBH, which not only provides a best-fit point
estimate but also allows for drawing waveform samples. We do so by
modeling posterior probability distributions of the model's NR-calibration
parameters, obtained from 441 NR simulations of aligned-spin BBHs
produced with the Spectral Einstein code of the
Simulating eXtreme Spacetimes (SXS) Collaboration~\cite{Boyle:2019kee,
	Mroue:2013xna, Hemberger:2013hsa, Scheel:2014ina, Lovelace:2014twa,
	Chu:2015kft, Blackman:2015pia, LIGOScientific:2016sjg,
	LIGOScientific:2016kms, Bohe:2016gbl, Lovelace:2016uwp,
	Blackman:2017dfb, Varma:2018mmi, Varma:2019csw, Yoo:2022erv}, and
one produced with the \texttt{Einstein Toolkit}~\cite{EinsteinToolkit:2024_05}.
We interpolate the probability distributions across parameter space using mixture density networks
(MDNs)~\cite{astonpr373}, a neural-network architecture suited for
modeling and predicting probability distributions over continuous
variables. Using these probability distributions as
priors, we sample the NR-calibration parameters
together with the standard source parameters during inference,
effectively marginalizing over the NR-calibration uncertainty in the model.

\section{\seob~ waveform model}
We use geometric units $G = 1 = c$, and set $M = m_1 +m_2$ and $\nu = m_1 m_2/M^2$, where $m_1$ and $m_2$ are the BH's masses.
In the EOB formalism, the binary's conservative dynamics is described by the EOB Hamiltonian $H_{\rm EOB} = M\, \sqrt{1 + 2 \nu (\hat{H}_{\rm eff} -1) }$, where $\hat{H}_{\rm eff} \equiv H_{\rm eff}/(\nu M)$ is the specific Hamiltonian of an effective test-body of reduced mass $\nu \,M$ moving in the (deformed) Kerr spacetime, being $0 \leq \nu \leq 1/4$ the deformation parameter. We also introduce the mass ratio $q = m_1/m_2 > 1$. We limit to aligned spins and introduce the combinations $\chi_{\rm{eff}}=(m_1 \chi_1 + m_2 \chi_2)/ M$ and $\chi_{\rm{a}}=(m_1 \chi_1 - m_2 \chi_2)/ M$.

The gravitational polarizations can be written as 
\begin{equation}
	h_+ - i h_\times = \sum_{\ell,m} {}_{-2} Y_{\ell m} (\varphi, \iota) h_{\ell m} (t),
\end{equation}
where ${}_{-2} Y_{\ell m} (\varphi, \iota)$ are the -2 spin-weighted spherical harmonics, with $\varphi$ and $\iota$ being the azimuthal and polar angles to the observer. The inspiral-merger-ringdown $(\ell, m)$ modes are given by $h_{\ell m} = h_{\ell m}^{\rm insp-plunge} \vartheta(t_{\text {peak }}^{22} - t) + h_{\ell m}^{\rm merg-RD} \vartheta(t - t_{\text {peak }}^{22})$, where $\vartheta$ is the Heaviside step function and $t_{\text {peak }}^{22}$ is the peak time of the $(2,2)$-mode amplitude.
The inspiral-plunge modes are based on a factorization of the PN GW modes~\cite{Damour:2008gu, Damour:2007xr, Pan:2010hz, Damour:2007yf}, evaluated on the dynamics obtained from the EOB equations of motion \cite{Buonanno:2000ef, Pan:2011gk}.
For the merger-ringdown part of the waveform, we use a phenomenological ansatz \cite{Baker:2008mj,Damour:2014yha, Bohe:2016gbl, Cotesta:2018fcv, Pompili:2023tna}, informed by NR and BH perturbation theory.

The accuracy of EOB waveforms is improved through calibration to NR simulations. For the inspiral-plunge stage, this is achieved by introducing higher-order PN terms in the EOB Hamiltonian, whose coefficients are tuned to NR, and fitting the merger time (i.e., $t_{\text {peak }}^{22}$) to NR. The \seob~model~\cite{Pompili:2023tna} employs three calibration parameters
$\boldsymbol{\theta} \equiv (\Delta t_{\rm NR}, a_6,d_{\rm SO})$. Here, $\Delta t_{\rm NR}$ is defined by $t_{\text {peak }}^{22}=t_{\mathrm{ISCO}}+\Delta t_{\mathrm{NR}}$, where $t_{\rm{ISCO}}$ is the time at which $r = r_{\rm{ISCO}}$, with $r_{\rm{ISCO}}$ the radius of the Kerr ISCO~\cite{Bardeen:1972fi} with the mass and spin of the remnant BH~\cite{Jimenez-Forteza:2016oae, Hofmann:2016yih}. The parameter $a_6$ is a 5PN nonspinning coefficient and $d_{\rm SO}$ is a 4.5PN spin-orbit coefficient in $H_{\rm{EOB}}$.
The \seob~model is calibrated to 442 aligned-spin NR waveforms covering mass ratios $q$ from $1$ to $20$ in the nonspinning limit, and  spins ranging from $-0.998 \leq \chi_i \leq 0.998$ for $q=1$ to $-0.5 \leq \chi_1 \leq 0.5,~\chi_2=0$ for $q=15$ (see Fig. 2 in Ref.~\cite{Pompili:2023tna}).
In this work, we focus on the inspiral-merger calibration parameters, as they have the largest impact on most signals, except those dominated by the ringdown stage. By modifying the orbital dynamics, these parameters provide a coherent correction to all waveform modes. Additionally, we focus on the spin sector since the nonspinning sector shows significantly higher accuracy~\cite{Pompili:2023tna}. Thus, we consider $\boldsymbol{\theta} = (\Delta t_{\rm NR}, d_{\rm SO})$.

\section{Probabilistic waveform model}
\label{sec:prob}
Waveform accuracy is often quantified by the mismatch $\mathcal{M}$ (or unfaithfulness), defined as $1$ minus the overlap between the normalized waveforms, maximized over a relative time and phase shift:
\begin{equation}
	\label{eq:mismatch}
	\mathcal{M}= 1 - \max _{\varphi_0, t_0} \frac{\left(h_1 \mid h_2\right)}{\sqrt{\left(h_1 \mid h_1\right)\left(h_2 \mid h_2\right)}}.
\end{equation}
The overlap is a noise-weighted inner product \cite{Sathyaprakash:1991mt, Finn:1992xs}
\begin{equation}
	\left(h_1 \mid h_2\right) \equiv 4 \operatorname{Re} \int_{f_l}^{f_h} df \tilde{h}_1(f) \tilde{h}_2^*(f) / S_n(f),
\end{equation}
where $\tilde h(f)$ is the Fourier transform of the time-domain signal, the ${}^*$ superscript indicates complex conjugation, and $S_n(f)$ is the power spectral density (PSD) of the detector noise, which we assume to be the zero-detuned high-power noise PSD of Advanced LIGO~\cite{Barsotti:2018}.~\footnote{Note that, while Eq.~\eqref{eq:mismatch} depends on the specific PSD used, both the mismatch against NR and the associated calibration errors do not depend strongly on the shape of the PSD. This was quantified in Appendix E of~\cite{Pompili:2023tna}, where a model calibrated against NR using the advanced LIGO PSD was compared to NR waveforms using the ET, CE and a flat PSD, yielding very comparable results.}
The optimal SNR of a signal $h$ is $\mathrm{SNR} = \sqrt{\left(h \mid h\right)}$.

The \seob~calibration procedure is outlined in Refs.~\cite{Bohe:2016gbl, Pompili:2023tna}. It essentially consists of determining values of the calibration parameters $\boldsymbol{\theta}$ that minimize a combination of the mismatch and the difference in merger time $\delta t_{\rm{merger}}$ (defined as the peak of the $(2,2)$-mode amplitude) between EOB and NR waveforms with the same physical parameters $\boldsymbol{\Lambda} \equiv(q, \chi_1, \chi_2)$.
This is carried out in a Bayesian fashion using the \texttt{Bilby}~\cite{Ashton:2018jfp} package, and the \texttt{pySEOBNR} code~\cite{Mihaylov:2023bkc}.
One defines a likelihood function of the form
\begin{equation}
	\label{eq:calib_likelihood}
	p\left(\boldsymbol{\Lambda}^{\mathrm{NR}} \mid \boldsymbol{\theta}\right) \propto \exp \left[-\frac{1}{2}\left(\frac{\mathcal{M}_{\max }(\boldsymbol{\theta})}{\sigma_{\mathcal{M}}}\right)^2-\frac{1}{2}\left(\frac{\delta t_{\text {merger }}(\boldsymbol{\theta})}{\sigma_t}\right)^2\right],
\end{equation}
where ${\mathcal{M}}_{\max}(\boldsymbol{\theta})$ is the maximum (2,2)-mode mismatch between EOB and NR waveforms over the total mass range $10 M \leq M_{\odot} \leq 200 M$. $\sigma_{\mathcal{M}}$ is chosen to be $10^{-3}$, and $\sigma_{t}=5 M$, these being conservative estimates of NR errors~\cite{Boyle:2019kee, Pompili:2023tna}.
A posterior distribution $p\left(\boldsymbol{\theta} \mid \boldsymbol{\Lambda}\right)$ is obtained for each NR configuration with parameters $\boldsymbol{\Lambda}$ using nested sampling~\cite{Skilling:2006gxv, Williams:2021qyt}, assuming uniform priors on $\boldsymbol{\theta}$. After a postprocessing step to select only one mode of each calibration posterior, in \seob~ one fits a point measure $\boldsymbol{\theta}(\boldsymbol{\Lambda})$ for each NR simulation, like the median of the distribution, over the \boldsymbol{$\Lambda$} parameter space, using least-square hierarchical fits. We use the final calibration posteriors from Ref.~\cite{Pompili:2023tna}.

In this work, we fit the \emph{full} posterior distributions of the calibration parameters $p\left(\boldsymbol{\theta} \mid \boldsymbol{\Lambda}\right)$, instead of point estimates $\boldsymbol{\theta}(\boldsymbol{\Lambda})$. This provides a point measure from, e.g., the mean of the posterior, while also giving access to parameter uncertainty. It can be interpreted as a probabilistic waveform model~\cite{Doctor:2017csx, Williams:2019vub, Khan:2024whs} by drawing samples of the calibration parameters. We model the calibration posteriors $p\left(\boldsymbol{\theta} \mid \boldsymbol{\Lambda}\right)$ as a mixture of Gaussians depending on physical parameters, using MDNs.
MDNs offer a versatile alternative to standard, single-output, GPR approaches~\cite{Khan:2024whs}, allowing for the modeling of not only the uncertainty in individual parameters, but also the correlations between different NR-calibration parameters. We define
\begin{equation}
	p({\boldsymbol{\theta}} \mid \boldsymbol{\Lambda}) \approx \sum_{k=1}^K \pi_k(\boldsymbol{\Lambda}) \mathcal{N}\left(\boldsymbol{\theta} \mid \mu_k(\boldsymbol{\Lambda}), \Sigma_k(\boldsymbol{\Lambda})\right),
\end{equation}
where $\mathcal{N}\left(\boldsymbol{\theta} \mid \mu, \Sigma\right)$ are multivariate Gaussian distributions with mean $\mu$ and covariance matrix $\Sigma$, e.g.,
$\mathcal{N}(\boldsymbol{\theta} \mid {\mu}, {\Sigma}) \propto \exp \left[-\frac{1}{2}(\boldsymbol{\theta}-{\mu})^T {\Sigma}^{-1}(\boldsymbol{\theta}-{\mu})\right]$,
and $\pi_k(\boldsymbol{\Lambda})$ are the weights of the $K$ Gaussians in the mixture.
The weights, means and covariance matrices are modeled with neural networks using the \texttt{PyTorch} framework~\cite{Paszke:2019xhz}.

To improve the extrapolation behavior of the model, we fit the residuals after subtracting the least-square fits of Ref.~\cite{Pompili:2023tna}, which we denote $\boldsymbol{\delta \theta} = (\delta \Delta t_{\rm{NR}}, \delta d_{\rm{SO}})$.
Specifically, we train the MDN to model the NR-calibration posterior residuals, denoted as $p(\boldsymbol{\delta \theta} \mid \boldsymbol{\Lambda})$.
Details on the neural network architecture and the sanity checks that we performed are provided in the Appendix~\ref{app:network}. Figure~\ref{fig:waveform} shows an \seob~waveform with its uncertainty estimate, from different calibration posterior samples, for $\boldsymbol{\Lambda} \equiv (q, \chi_1, \chi_2) = (3.0, 0.45, 0.45)$, compared against the \texttt{NRHybSur3dq8} model.

\begin{figure}
	\vspace{-10pt}
	\includegraphics[width=0.50\textwidth]{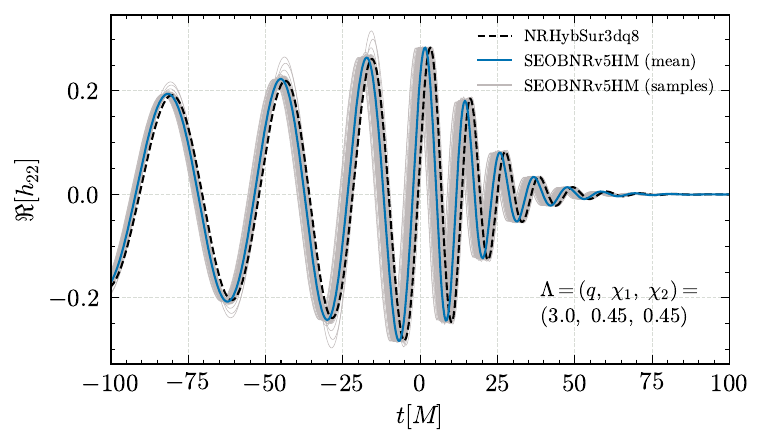}
	\caption{An \seob~waveform with its uncertainty estimate, compared against the \texttt{NRHybSur3dq8} model after a low-frequency alignment. The waveforms start at a frequency of $20~\rm{Hz}$ for a binary with total mass $M=60~M_{\odot}$.
	We show the \seob~waveform using as NR-calibration parameters $\boldsymbol{\theta}$ the mean of the MDN fit of the calibration posterior $p\left(\boldsymbol{\theta} \mid \boldsymbol{\Lambda}\right)$, while the uncertainty corresponds to 100 samples from the same distribution.
	The time $t = 0$ corresponds to the peak of the $(2,2)$ mode of the NR waveform.
	}
	\label{fig:waveform}
	\vspace{-10pt}
\end{figure}

We marginalize over modeling uncertainties in PE by sampling both the source parameters and corrections to the NR-calibration parameters $\boldsymbol{\delta\theta}$, using $p\left(\boldsymbol{\delta \theta} \mid \boldsymbol{\Lambda}\right)$ as conditional priors.
As in Ref.~\cite{Owen:2023mid} we also compare these results to those obtained using uniform priors on $\boldsymbol{\delta \theta}$. We choose a range within $[-50, 50]$ for both $\delta \Delta t_{\rm{NR}}$ and $\delta d_{\rm{SO}}$, slightly broader than the typical range of NR-calibration posteriors.

\section{Results} We now assess the impact of marginalizing over the NR-calibration uncertainty in the parameter inference of loud ``golden'' BBH systems. The GW strain of quasicircular, aligned-spin BBHs is characterized by 11 parameters $\boldsymbol{\lambda}=\left\{\mathcal{M}_c, 1/q, \chi_1, \chi_2, d_{\mathrm{L}}, \alpha, \delta, \iota, \varphi, \psi, t_c \right\}$.
We parametrize the masses in terms of the chirp mass $\mathcal{M}_c = (m_1 m_2)^{3/5}/M^{1/5}$ and inverse mass ratio $1/q$. The position of the binary is described by its luminosity distance, $d_L$, and the coordinates on the plane of the sky, $(\alpha, \delta)$. The orientation of the binary is described by the polar angle, $\iota$, and the azimuthal angle, $\varphi$, to the observer in the source frame~\cite{Schmidt:2017btt} at the reference frequency, which we always set to $f_{\rm ref}=20\rm~Hz$. Finally, the relative contribution of the two gravitational polarizations, $h_+(t)$ and $h_{\times}(t)$, is described by the polarization angle, $\psi$, while the reference for the time is given by the coalescence time, $t_c$.

To simulate and analyze the GW signals we use the \texttt{parallel Bilby} package~\cite{Ashton:2018jfp, Smith:2019ucc, Romero-Shaw:2020owr} and the nested sampler \texttt{Dynesty}~\cite{Speagle:2019ivv}. We employ standard priors for all the parameters~\cite{Romero-Shaw:2020owr}. We consider four detector network configurations: a O4 network (corresponding to the current fourth observing run) with two advanced LIGO at O4 sensitivity~\cite{LIGO:NoiseCurves} and one advanced Virgo detector at design sensitivity~\cite{KAGRA:2013rdx}; two upgraded LIGO networks with O5 and A\# sensitivities~\cite{LIGO:Asharp} alongside advanced Virgo; and one ET-A\# network with LIGO detectors at A\# sensitivities combined with a 10 km ET at Virgo.
The low-frequency cutoff varies across these networks: $20~\rm{Hz}$ for O4, $15~\rm{Hz}$ for O5, $10~\rm{Hz}$ for the A\# and the ET-A\# network.~\footnote{It would be more realistic to consider the ET starting from a lower frequency of $3-5~\rm{Hz}$. However, the accuracy of current long NR waveforms is likely insufficient for this, as their accuracy during the inspiral phase is compromised by accumulated phase errors. This also makes it difficult to quantify the accuracy of hybridized NR surrogate waveforms, which we use to simulate signals, for such long durations~\cite{Varma:2018mmi}.}
Additional details about the detector network are given in Appendix~\ref{app:detectors}.

For each detector network, we simulate two BBH signals in zero-noise using the \texttt{NRHybSur3dq8} model, selecting configurations with unequal masses ($q=3$) and either positive ($\chi_1 = \chi_2 = 0.45$) or negative ($\chi_1 = \chi_2 = - 0.45$) spins. These configurations fall within the model's NR-calibration region but are still relatively challenging, so mismodeling errors may be significant at high SNR. We set the detector-frame total mass of the binary to $M = 60 M_{\odot}$, which gives $\mathcal{M}_c \simeq 21.976$ for $q=3$. Common parameters for both systems include: inclination angle $\iota = 0.438$, azimuthal angle $\varphi = 0$, sky location $\alpha = 1.827$, $\delta = -1.252$ (radians), luminosity distance $d_L = 600.0 \mathrm{Mpc}$, polarization angle $\psi = 2.562$ and GPS time at the geocenter $1126259462.409$ s. For the negative (positive)-spin configuration, the network optimal SNR ranges from $39.3$ ($51.8$) in the O4 network, to $366.7$ ($439.2$) in the ET-A\# network. While the calibration procedure described in Sec.~\ref{sec:prob} only uses the $(2,2)$ mode, as this is sufficient to calibrate the EOB orbital dynamics, we now include higher-order modes in both the injected signal and the recovery templates, using all modes provided by default in the respective models.

For each of the 8 configurations, we perform three recoveries. In the first, we use the default \seob~model and sample only the standard source parameters. In the other two, we account for NR-calibration uncertainty by sampling over corrections to the NR-calibration parameters $\boldsymbol{\delta\theta}$ with either $p\left(\boldsymbol{\delta \theta} \mid \boldsymbol{\Lambda}\right)$ or uniform priors, as previously described.

\begin{figure}
	\vspace{-10pt}
	\includegraphics[width=0.99\columnwidth]{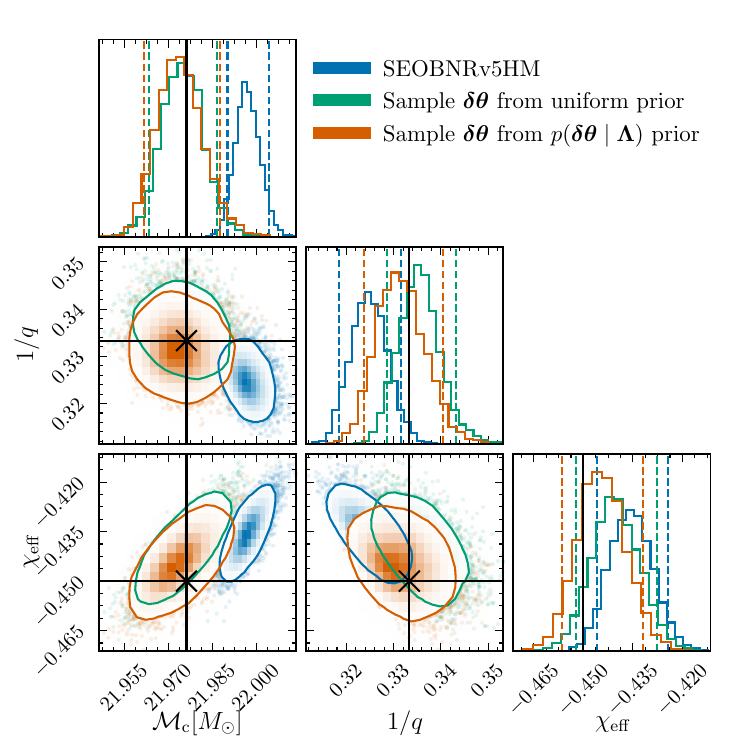}
	\caption{Marginalized posterior distributions for the chirp mass, inverse mass ratio, and effective spin in the ET-A\# detector network. Parameter estimation is performed by injecting a \texttt{NRHybSur3dq8} signal, with an SNR of $366.72$, and recovering it using three versions of \seob: one sampling only the standard source parameters, and the others including corrections to the NR-calibration parameters $\boldsymbol{\delta\theta}$. For $\boldsymbol{\delta\theta}$, both uniform priors and $p\left(\boldsymbol{\delta \theta} \mid \boldsymbol{\Lambda}\right)$ priors, reflecting NR-calibration uncertainty, are used. In the latter case, we accurately recover the injected values within the 90\% credible region, indicated by dashed vertical lines in the 1D marginalized posteriors and contours in the 2D marginalized posteriors.
	}
	\label{fig:corner}
	\vspace{-10pt}
\end{figure}

\begin{figure*}[t]
	\vspace{-10pt}
	\begin{minipage}{0.48\textwidth}
		{
			\stackinset{l}{0cm}{t}{0cm}{\textbf{a}}
			{\raggedright \includegraphics[width=\textwidth]{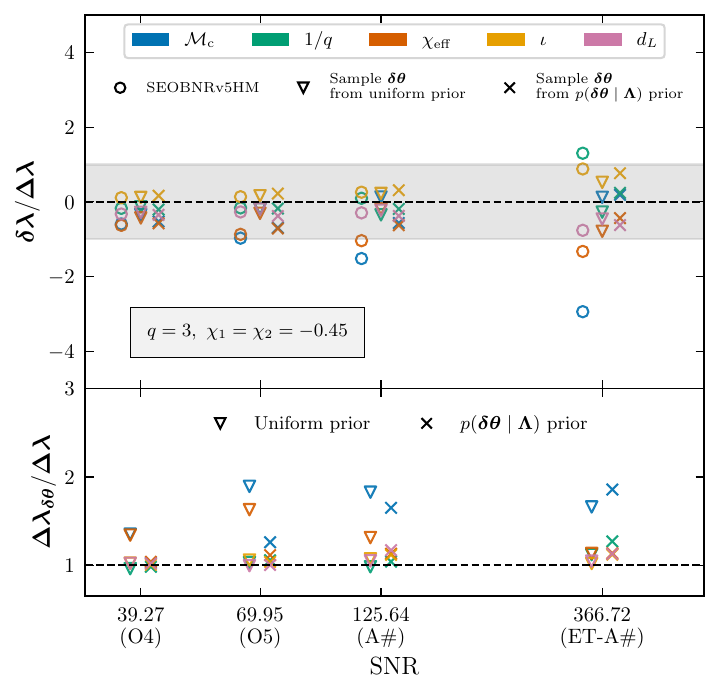}}}
	\end{minipage}
	\hfill
	\begin{minipage}{0.48\textwidth}
		{
			\stackinset{l}{0cm}{t}{0cm}{\textbf{b}}
			{\raggedright
				\includegraphics[width=\textwidth]{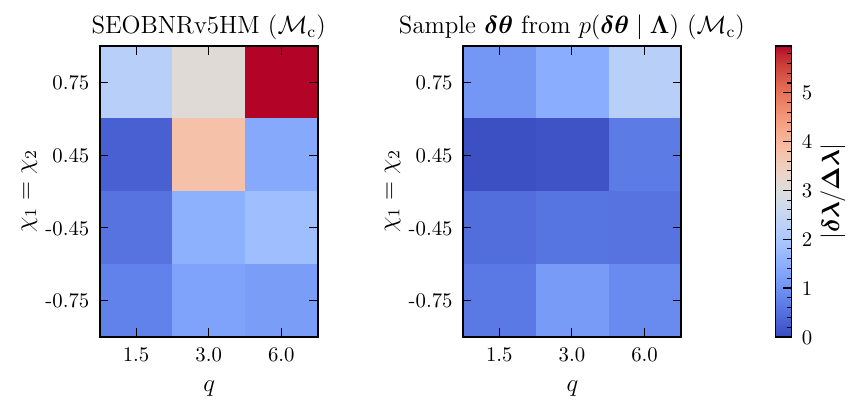}
			}}
		\includegraphics[width=\textwidth]{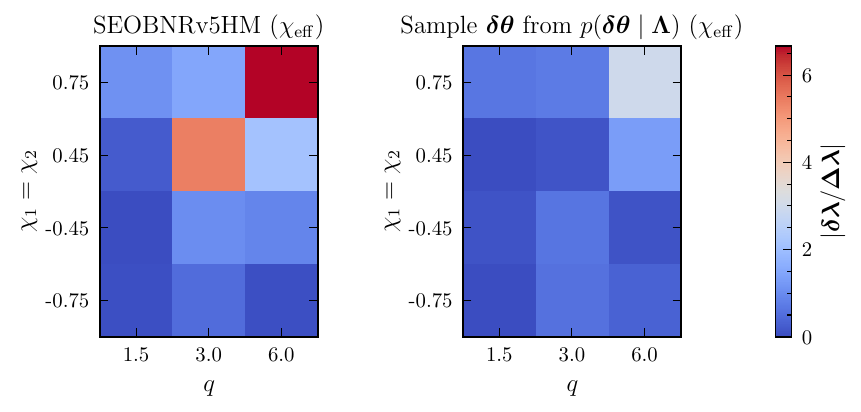}
	\end{minipage}
	\caption{(a) \emph{Top}: ratio of systematic bias to statistical errors ($\boldsymbol{\delta \lambda} / \boldsymbol{\Delta \lambda}$) for the chirp mass, inverse mass ratio, effective spin, inclination, and luminosity distance, as a function of the network SNR, for upcoming runs, O4, O5, A\#, and next-generation ground-based detector networks (ET-A\#). Parameter estimation is performed by injecting a \texttt{NRHybSur3dq8} signal, and recovering it with three versions of \seob: one sampling only the standard parameters, and the others including corrections to the NR-calibration parameters $\boldsymbol{\delta\theta}$. For $\boldsymbol{\delta\theta}$, both uniform priors and $p\left(\boldsymbol{\delta \theta} \mid \boldsymbol{\Lambda}\right)$ priors, reflecting NR-calibration uncertainty, are used. \emph{Bottom}: ratio of statistical errors ($\boldsymbol{\Delta \lambda}$) with and without corrections to the NR-calibration parameters $\boldsymbol{\delta\theta}$, using different priors.
		(b) Ratio of systematic bias to statistical errors ($|\boldsymbol{\delta \lambda} / \boldsymbol{\Delta \lambda}|$) across parameter space, for the chirp mass (top row), and effective spin (bottom row), in the A\# detector network.
	}
	\label{fig:bias_snr}
	\vspace{-10pt}
\end{figure*}

Figure~\ref{fig:corner} shows marginalized posterior distributions for the chirp mass, inverse mass ratio, and effective spin, for the negative-spin configuration in the ET-A\# detector network (SNR$=366.7$). The \seob~recovery shows biases in $\mathcal{M}_{\rm{c}}$, $1/q$, $\chi_{\rm{eff}}$, with the injected value falling outside of the 90\% credible region, indicated by dashed vertical lines in the 1D marginalized posteriors and contours in the 2D marginalized posteriors.
Sampling over $\boldsymbol{\delta \theta}$ with uniform priors mitigates these biases, bringing the injected parameters at the edge of the 90\% contours.
This improvement is partially due to the increase in statistical error, though all parameters also peak closer to the injected values. Using $p\left(\boldsymbol{\delta \theta} \mid \boldsymbol{\Lambda}\right)$ priors for $\boldsymbol{\delta \theta}$ further improves the recovery of $1/q$ and $\chi_{\rm{eff}}$, with the injected values now falling well within the 2D posteriors.
The increase in statistical error and decrease in systematic error when accounting for waveform uncertainties is qualitatively consistent with the earlier results of Refs.~\cite{Moore:2014pda, Gair:2015nga, Moore:2015sza} and with the PN model of Ref.~\cite{Owen:2023mid}. Biases arise because the template model can provide a better fit to the signal when the parameters are not the true ones. By improving the match through adjustments in ``nuisance'' parameters like $\boldsymbol{\delta \theta}$, the need for shifts in the physical parameters is reduced. The increase in statistical error is due to additional correlations introduced by the nuisance parameters.

The left panel of Fig.~\ref{fig:bias_snr} show the ratio of systematic bias to statistical errors ($\boldsymbol{\delta \lambda} / \boldsymbol{\Delta \lambda}$) for the same set of parameters, as well as inclination and luminosity distance, as a function of the SNR across the four detector networks. We focus on the negative-spin configuration; in the positive-spin case, significant biases still appear in the ET-A\# network, and we discuss possible motivations in the Appendix~\ref{app:higher-modes}.
We take $\boldsymbol{\delta \lambda}$ to be the difference between the injected parameters and the median of the 1D marginalized posteriors, and $\boldsymbol{\Delta \lambda}$ to be half of the width of the 90\% 1D credible interval.
At SNRs of $\sim100$ and higher, the original model exhibits biases in masses and spins (e.g., $|\boldsymbol{\delta \lambda} / \boldsymbol{\Delta \lambda}| > 1$) which are substantially mitigated when accounting for NR-calibration uncertainties, with all cases having $|\boldsymbol{\delta \lambda / \Delta} \boldsymbol{\lambda}| \lesssim 1$ when including corrections to the NR-calibration parameters $\boldsymbol{\delta\theta}$.
To understand how much the reduction of $|\boldsymbol{\delta \lambda} / \boldsymbol{\Delta \lambda}|$ is due to an increase in statistical errors, we show in the bottom panel the ratio of $\boldsymbol{\Delta \lambda}$ with and without $\boldsymbol{\delta\theta}$ corrections. For both prior choices, there is an increase of a factor of a few in statistical errors, mainly for masses and spins, with a marginal increase for distance and inclination. MDN-informed priors lead to a smaller increase in statistical errors, especially at lower SNRs where the default \seob~model is unbiased. The increase in statistical error can be below the $\boldsymbol{\delta \lambda} / \boldsymbol{\Delta \lambda}$ ratio of the default~\seob~model, even when models with uncertainty corrections achieve $\boldsymbol{\delta \lambda} / \boldsymbol{\Delta \lambda} \lesssim 1$, indicating a shift of the recovered parameters toward the injected values.

To assess the effectiveness of our method across parameter space, we also explore a larger set of $12$ configurations with mass ratios $q \in [1.5, 3.0, 6.0]$ and equal spins $\chi_i \in [-0.75, -0.45, 0.45, 0.75]$, fixing the chirp mass to $\mathcal{M}_c \simeq 21.976$. For simplicity, we limit our analysis to the A\# detector network. At a fixed luminosity distance, the SNR of the injected signal decreases with increasing mass ratio and increases with higher spins, ranging from $88.0$ for the configuration with $q=6$ and $\chi_i = - 0.75$, to $187.8$ for $q=1.5$ and $\chi_i = 0.75$.
In this case, we recover \texttt{NRHybSur3dq8} signals using \seob, either sampling only the standard source parameters, or including corrections to the NR-calibration parameters $\boldsymbol{\delta\theta}$ with $p\left(\boldsymbol{\delta \theta} \mid \boldsymbol{\Lambda}\right)$ priors.
The right panels of Fig.~\ref{fig:bias_snr} show the ratio of systematic bias to statistical errors ($|\boldsymbol{\delta \lambda} / \boldsymbol{\Delta \lambda}|$) across the parameter space, for the chirp mass and $\chi_{\rm{eff}}$. The left columns presents the default \seob~recovery, while the right columns shows results sampling on corrections to the NR-calibration parameters $\boldsymbol{\delta\theta}$ with $p\left(\boldsymbol{\delta \theta} \mid \boldsymbol{\Lambda}\right)$ priors. Including corrections to the NR-calibration parameters generally reduces biases across the parameter space, especially for the most challenging configurations with large positive spins and unequal mass ratios.

While biases are generally reduced, they are not always completely eliminated, such as for the case with $q = 6$ and $\chi_i = 0.75$. When considering an approximate model $H(\boldsymbol{\Lambda})$ and the ``true'' waveform $h_{\rm{true}}(\boldsymbol{\Lambda})$, we only have an approximate relation $h_{\mathrm{true}} (\boldsymbol{\Lambda}) \simeq H(\boldsymbol{\Lambda}, \boldsymbol{\theta}(\boldsymbol{\Lambda}) + \boldsymbol{\delta \theta}(\boldsymbol{\Lambda}))$. The effectiveness of the $\boldsymbol{\delta \theta}$ correction in reducing the difference between the approximate and true waveforms depends crucially on the choice of the calibration parameters $\boldsymbol{\theta}$. Additional corrections, such as higher-order terms in the EOB Hamiltonian and energy flux, or corrections to higher modes, may be relevant depending on the source's properties and SNR.
The $\boldsymbol{\delta \theta}$ corrections we focus on account for the inevitable errors in the fits to NR, as well as the fact that NR is only an approximation of the true signal. While this may represent only a partial source of error, it is inherently linked to the construction of all inspiral-merger-ringdown models calibrated to NR, and must be accounted for.

Despite adding extra parameters, the computational cost of recoveries including $\boldsymbol{\delta \theta}$ is lower than for the default model, due to fewer likelihood evaluations needed for convergence. This is especially true with MDN-based priors on $\boldsymbol{\delta \theta}$, with $\sim 30\%$ fewer likelihood evaluations required.

\section{Conclusions} In this work, we introduced a method to mitigate
waveform systematic errors, by incorporating and marginalizing over
waveform uncertainty estimates. We modeled the latter as probability distributions
for the \seob~\cite{Pompili:2023tna} NR-calibration parameters using
MDNs. We demonstrated its effectiveness in significantly reducing
systematic errors in the parameter inference of loud ``golden'' BBH
systems observed with upcoming and XG detectors, at the cost of a
slight increase in statistical uncertainty.

This method addresses uncertainties within the NR-calibration
region of the model, but does not necessarily lead to increasing
error estimates in regions of the parameter space lacking training data.
However, model differences can be
large outside their NR-calibration region~\cite{Dhani:2024jja}.
To account for these extrapolation uncertainties, one could introduce error estimates based
on model-to-model differences rather than relying solely on NR
data. Additional nuisance parameters, such as higher-order PN terms or
merger-ringdown fitted coefficients, may also be necessary for more extreme
binary configurations.
Our method addresses a specific source of error,
related to the calibration of waveform models to NR, but does not account,
e.g., for missing physical effects. Therefore, incorporating all relevant physical
effects into waveform models remains essential. However, our results also indicate
that this source of error would be significant in the near future, potentially as early as O5,
and must be accounted for.
For events from previous LVK observing runs, we expect that our uncertainty estimates would not yield significant differences, given the relatively low SNR. Reanalyses of the events GW150914 and GW190412 using the publicly released strain data~\cite{LIGOScientific:2019lzm, KAGRA:2023pio} show a maximum Jensen-Shannon divergence across different parameters of $0.004$ bits between recoveries with and without uncertainty estimates, comparable to the statistical uncertainty of stochastic sampling algorithms~\cite{Romero-Shaw:2020owr}.

An advantage of our method is that, by using the native time-domain \seob~model,
it can already be applied to its spin-precessing and eccentric BBH extensions,
as well as systems containing NSs, to account for NR-calibration uncertainties in
the quasicircular aligned-spin BBH sector common to all the
models. Once EOB models are calibrated to eccentric and
spin-precessing NR simulations, our method can be extended to handle
specific uncertainties related to these features.
Our results with uniform priors on $\boldsymbol{\delta \theta}$ also suggest that PE for spin-precessing and eccentric
BBHs could be improved even without detailed calibration to NR, provided that
reasonable priors can be set on appropriate calibration parameters.
Our method could also enhance the robustness of tests of GR based on EOB
waveforms~\cite{Brito:2018rfr, Ghosh:2021mrv, Maggio:2022hre,
	Toubiana:2023cwr, Mehta:2022pcn}, and may help mitigate false GR
deviation caused by waveform systematics~\cite{Maggio:2022hre, Toubiana:2023cwr}.

\medskip

\section*{Acknowledgments} We are grateful to Arnab Dhani, Héctor Estellés, Jonathan Gair, Elise Sänger, Hector O. Silva and Alexandre Toubiana for valuable discussions.
We also thank Cecilio Garcia-Quiros and Sebastian Khan for useful comments on the manuscript.
The computational work for this manuscript was carried out on the \texttt{Hypatia} computer cluster at the Max Planck Institute for Gravitational Physics.

This study is similar to the recent work by Bachhar \emph{et al.}~\cite{Bachhar:2024olc}, as both introduce a method for modeling waveform uncertainty from calibrating EOB models against NR data, and employ it in a Bayesian inference study. However, they differ significantly in how the uncertainty models are constructed and incorporated into PE.

M.P. was supported by NSF Grants AST-2407453 and PHY-2512902.
The material presented in this paper is based upon work supported by National
Science Foundation's (NSF) LIGO Laboratory, which is a major facility fully
funded by the NSF.
This research has made use of data or software obtained from the Gravitational
Wave Open Science Center (\href{gwosc.org}{gwosc.org}), a service of LIGO
Laboratory, the LIGO Scientific Collaboration, the Virgo Collaboration, and
KAGRA. LIGO Laboratory and Advanced LIGO are funded by the United States
National Science Foundation (NSF) as well as the Science and Technology
Facilities Council (STFC) of the United Kingdom, the Max-Planck-Society (MPS),
and the State of Niedersachsen/Germany for support of the construction of
Advanced LIGO and construction and operation of the GEO600 detector.
Additional support for Advanced LIGO was provided by the Australian Research
Council. Virgo is funded, through the European Gravitational Observatory (EGO),
by the French Centre National de Recherche Scientifique (CNRS), the Italian
Istituto Nazionale di Fisica Nucleare (INFN) and the Dutch Nikhef, with
contributions by institutions from Belgium, Germany, Greece, Hungary, Ireland,
Japan, Monaco, Poland, Portugal, Spain. KAGRA is supported by Ministry of
Education, Culture, Sports, Science and Technology (MEXT), Japan Society for
the Promotion of Science (JSPS) in Japan; National Research Foundation (NRF)
and Ministry of Science and ICT (MSIT) in Korea; Academia Sinica (AS) and
National Science and Technology Council (NSTC) in Taiwan.

\section*{Data availability}

The data that support the findings of this article are not publicly available. The data are available from the authors upon reasonable request.

\bibliography{bibliography}

\clearpage

\appendix

\section{Detector networks}
\label{app:detectors}
\begin{figure}
	\includegraphics[width=0.50\textwidth]{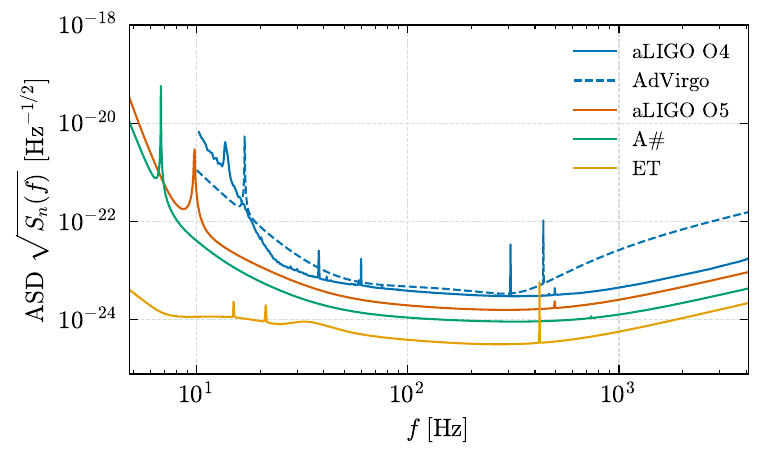}
	\caption{
		The amplitude-spectral-density curves of the various detectors used in this paper. The curves labeled by aLIGO O4 and AdVirgo denote the sensitivity of the LIGO detectors during the fourth observing run, and the design sensitivity of the Virgo detector, respectively. The aLIGO O5 and A\# curves refer to the design (A+) and upgraded sensitivity of the LIGO detectors, respectively.}
	\label{fig:plot_asd}
\end{figure}

Current GW detectors are projected to reach their design sensitivity within the next few years during the fifth observing run (O5) and will continue to operate until the end of the decade~\cite{KAGRA:2013rdx}. Following this period, significant upgrades are anticipated, with the expectation that detectors might remain operational until the advent of XG detectors. In this work, we consider four GW ground-based detector networks, consisting of current detectors at their design and upgraded sensitivities, and a configuration where upgraded current detectors work in conjunction with an XG detector. These are listed below:
\begin{itemize}
	\item O4 network: This network comprises the advanced LIGO detectors at Hanford and Livingston operating at O4 sensitivity, and the advanced Virgo detector operating at design sensitivity. We use the noise curves available in \texttt{Bilby}, specifically \texttt{aLIGO O4 high}~\cite{LIGO:NoiseCurves} and \texttt{AdV}~\cite{KAGRA:2013rdx}. The minimum frequency is set to $f_{\rm{low}} = 20~\rm{Hz}$.
	\item O5 network: This network includes the advanced LIGO detectors at Hanford and Livingston operating at design (A+) sensitivity, using the same noise curve as in~\cite{Dhani:2024jja}, along with the advanced Virgo detector at design sensitivity. The minimum frequency is set to $f_{\rm{low}} = 15~\rm{Hz}$.
	\item A\# network: This network consists of the advanced LIGO detectors at Hanford and Livingston operating at A\# sensitivity, using the same noise curve as in~\cite{Dhani:2024jja}, and the advanced Virgo detector at design sensitivity. The minimum frequency is set to $f_{\rm{low}} = 10~\rm{Hz}$.
	\item ET-A\# network: This network includes a 10 km ET in Europe, at the Virgo location, using the same noise curve as in~\cite{Dhani:2024jja}, combined with LIGO detectors at Hanford and Livingston operating at A\# sensitivity. Although the ET is proposed to have a triangular configuration, the \texttt{bilby pipe} wrapper is limited to L-shaped interferometer configurations. Therefore, we assume the ET to be L-shaped as in Ref.~\cite{Dhani:2024jja}. This does not significantly impact our results as the shape of the interferometer is not critical to the conclusions discussed here~\cite{Branchesi:2023mws}. The minimum frequency at which we start integrating the likelihood is set to $f_{\rm{low}} = 10~\rm{Hz}$. While it would be more realistic to consider the ET starting from a lower frequency of $\sim 3-5~\rm{Hz}$, the accuracy of long NR simulations and of hybridized NR surrogate waveforms, which we use to simulate signals, is not guaranteed to be sufficient for such long signals~\cite{Varma:2018mmi}.
\end{itemize}
The amplitude-spectral-density (ASD) curves of the individual detectors are shown in Fig.~\ref{fig:plot_asd}. The details of the detector networks are further summarized in Table~\ref{tab:detector-networks}, together with the segment length and the network optimal SNR, for the binary configurations considered in this study. The maximum frequency is $f_{\rm{high}} = 1024~\rm{Hz}$ for all networks, which is above the ringdown frequency of the higher-order modes for all the systems considered in this work. The sampling frequency is then chosen to be $f_{\rm{s}}=2048~\rm{Hz}$. We determined optimal analysis settings with the \texttt{PE configurator} package \cite{peconfigurator}.

Proposed XG detector networks often include the ET operating in conjunction with one or more CE detectors. The sources we consider would have an SNR around $1500$ in such a network. For such high SNRs, the accuracy of current NR waveforms  may not be adequate~\cite{Purrer:2019jcp}, potentially complicating efforts to identify the origins of any biases. As a result, we do not explore these configurations.

\begin{table*}
	\begin{tabular}{c|c|c|c|c|c}
		\hline
		\hline
		Network & List of Interferometers and PSD                    & $f_\mathrm{low} [\mathrm{Hz}]$ & Segment length $[\rm{s}]$ &
		\multicolumn{2}{c}{Network SNR}                                                                                                                                                                  \\
		        &                                                    &                                &                           & \multicolumn{1}{c|}{$\chi_i > 0$} & \multicolumn{1}{c}{$\chi_i < 0$} \\
		\hline
		\hline
		O4      & H1, L1 (\texttt{aLIGO O4 high}), V1 (\texttt{AdV}) & 20                             & 8                         & 51.8                              & 39.3                             \\
		O5      & H1, L1 (\texttt{A+}), V1 (\texttt{AdV})            & 15                             & 8                         & 91.7                              & 70.0                             \\
		A\#     & H1, L1 (\texttt{A\#}), V1 (\texttt{AdV})           & 10                             & 16                        & 163.3                             & 125.6                            \\
		ET-A\#  & H1, L1 (\texttt{A\#}), ET (\texttt{ET-D})          & 10                             & 16                        & 439.2                             & 366.7                            \\
		\hline
		\hline
	\end{tabular}
	\caption{
		\label{tab:detector-networks}
		Detector networks used in this study. The networks are defined by the list of interferometers, with their PSD in parentheses. We also indicate the frequency $f_\mathrm{low}$ at which we start integrating the likelihood integral and the network SNR of the \texttt{NRHybSur3dq8} signals in these networks (see the main text for the parameters).
	}
\end{table*}

\section{Details of the MDN model and sanity checks}
\label{app:network}
This section provides technical details about the neural network architecture and summarizes some of the sanity checks that we performed to assess the robustness of our model.

We implement the MDN model using the \texttt{PyTorch} framework~\cite{Paszke:2019xhz}. Specifically, we construct the network using a \texttt{Sequential} container comprising multiple \texttt{Linear} layers. To introduce nonlinearity within the network layers, we apply a leaky ReLU (\texttt{LReLU}) activation function. The training process involves minimizing the following loss function with respect to the network weights $w$,
\begin{multline}
	L({w})=-\frac{1}{N} \sum_{n=1}^N
	\ln \left [\sum_{k=1}^K \pi_k\left(\boldsymbol{\Lambda}^{(n)}, {w}\right) \times \right . \\ \left . \mathcal{N}\left({\boldsymbol{\theta}}^{(n)} \mid \mu_k\left(\boldsymbol{\Lambda}^{(n)}, {w}\right), \Sigma_k\left(\boldsymbol{\Lambda}^{(n)}, {w}\right)\right)\right].
\end{multline}
Here, $\mathcal{N}\left({\boldsymbol{\theta}} \mid \mu, \Sigma\right)$ are multivariate Gaussian distributions with mean $\mu$ and covariance matrix $\Sigma$, e.g., $\mathcal{N}({\boldsymbol{\theta}} \mid {\mu}, {\Sigma}) \propto \exp \left[-\frac{1}{2}({\boldsymbol{\theta}}-{\mu})^T {\Sigma}^{-1}({\boldsymbol{\theta}}-{\mu})\right]$, $\pi_k$ are the weights of the $K$ Gaussians in the mixture, and $N$ is the number of NR-calibration posteriors used for training.
This loss function represents the mean of the negative log-posterior of the MDN~\cite{Kruse:2020}. We use full covariance matrices, which we parametrize by their Cholesky decomposition.

We use the \texttt{Adam} optimization algorithm for training the network. We initially determine a set of model hyper-parameters (number of Gaussians in the mixture, number of network layers, \texttt{n$\_$layers}, number of neurons per layer, \texttt{n$\_$neurons}, and number of training epochs, \texttt{n$\_$epochs}) with an 80-20 training-test split. These hyperparameters are further refined by performing waveform sanity checks against \texttt{NRHybSur3dq8} waveforms not included in the training set. Our final settings are \texttt{n$\_$layers} $= 6$, \texttt{n$\_$neurons} $= 96$, \texttt{n$\_$epochs} $= 5000$ and a single Gaussian component.

We transform the variables $\boldsymbol{\Lambda}$ from $(q, \chi_1, \chi_2)$ to $(\nu, \chi_{\rm{eff}}, \chi_{\rm{a}})$.
As mentioned in the main text, to improve the extrapolation behavior of the model, we fit the residuals of the calibration parameters $\boldsymbol{\theta}$ after subtracting the least-square fits of Ref.~\cite{Pompili:2023tna} (with the superscript $\rm{v5}$), which we denote $(\delta \Delta t_{\rm{NR}}, \delta d_{\rm{SO}}) \equiv \boldsymbol{\delta \theta}$. Test-particle ($\nu \rightarrow 0$) information is already enforced by the least-square fits, so we rescale the residual posteriors by $\nu$ to have the correction go to zero in that limit. In the small-spin limit, the posteriors are uninformative because the model is already well-calibrated, and the term containing $d_{\rm{SO}}$ in $H_{\rm{EOB}}$ is scaled by a $\chi_{\mathrm{eff}}$ factor~\cite{Khalil:2023kep}. To recover the nonspinning limit of the model, and improve its robustness for small-spin configurations, we taper the corrections with a $\tanh$ function
\begin{subequations}
	\begin{align}
		\Delta t_{\rm{NR}} & =\Delta t_{\rm{NR}}^{\rm{v5}}\left(\nu, \chi_{\mathrm{eff}}, \chi_{\mathrm{a}}\right)
		+\nu \tanh \left[\alpha\left(\chi_{\mathrm{eff}}^2+\chi_{\mathrm{a}}^2\right)\right] \Delta t_{\rm{NR}}^{\rm{MDN}}\left(\nu, \chi_{\mathrm{eff}}, \chi_{\mathrm{a}}\right) \nonumber                                                                                             \\
		                   & \equiv \Delta t_{\rm{NR}}^{\rm{v5}}\left(\nu, \chi_{\mathrm{eff}}, \chi_{\mathrm{a}}\right) + \delta \Delta t_{\rm{NR}}\,,                                                                                                                                  \\
		d_{\rm{SO}}        & =d_{\rm{SO}}^{\rm{v5}}\left(\nu, \chi_{\mathrm{eff}}, \chi_{\mathrm{a}}\right)+\nu \tanh \left[\alpha\left(\chi_{\mathrm{eff}}^2+\chi_{\mathrm{a}}^2\right)\right] d_{\rm{SO}}^{\rm{MDN}}\left(\nu, \chi_{\mathrm{eff}}, \chi_{\mathrm{a}}\right) \nonumber \\
		                   & \equiv d_{\rm{SO}}^{\rm{v5}}\left(\nu, \chi_{\mathrm{eff}}, \chi_{\mathrm{a}}\right) + \delta d_{\rm{SO}}.
	\end{align}
\end{subequations}
We choose $\alpha=10$ based on sanity checks against NR and \texttt{NRHybSur3dq8} waveforms not used in the training. Before performing the fit, we finally apply a \texttt{StandardScaler} transformation with \texttt{scikit-learn}~\cite{scikit-learn}, such that the data have a mean of zero and a standard deviation of one.

\begin{figure}
	\includegraphics[width=0.50\textwidth]{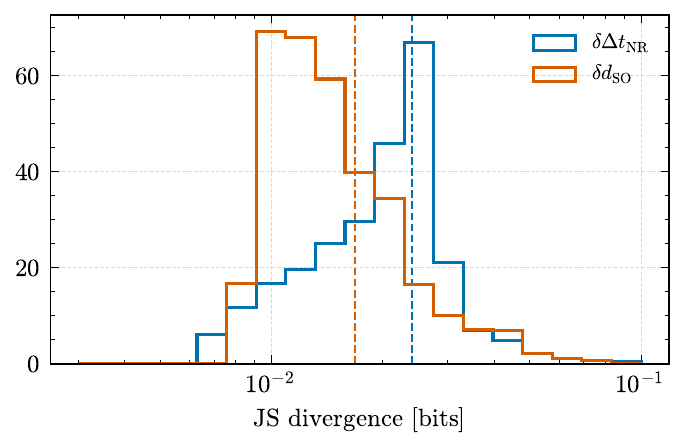}
	\caption{JSD between the 1D marginalized distributions of $(\delta \Delta t_{\rm{NR}}, \delta d_{\rm{SO}})$ for NR-calibration posteriors residuals, $p(\boldsymbol{\delta \theta} \mid \boldsymbol{\Lambda})$, and our MDN fit, across the 400 spinning NR simulations used to build the model. The vertical dashed lines indicate the median of the distributions. }
	\label{fig:jsd_plot}
\end{figure}

\begin{figure}
	\includegraphics[width=0.47\textwidth]{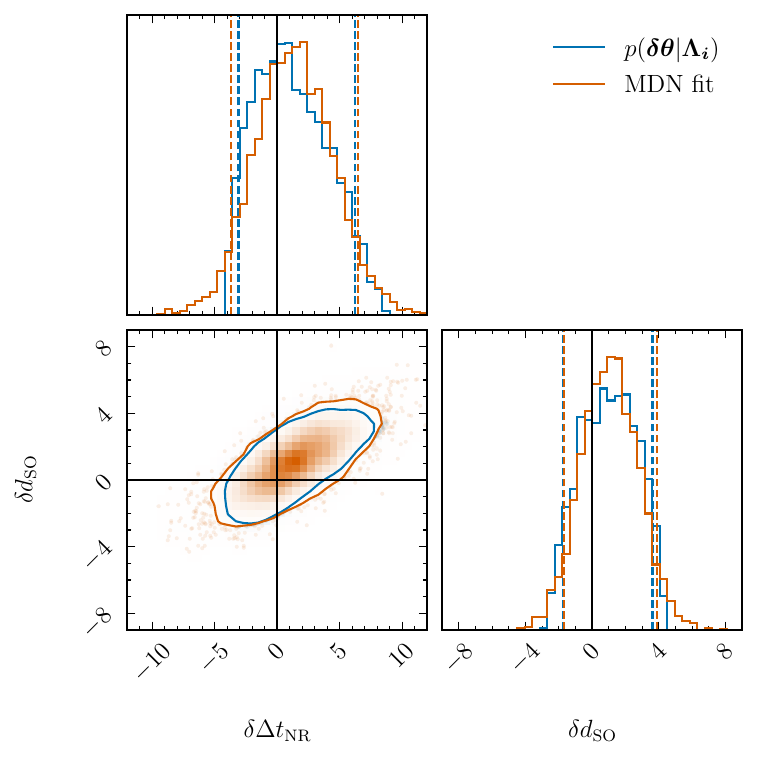}
	\caption{Comparison of the NR-calibration posterior residual $p(\boldsymbol{\delta \theta} \mid \boldsymbol{\Lambda})$ and the corresponding MDN fit, for the NR simulation \texttt{SXS:BBH:2158} with parameters $\boldsymbol{\Lambda_i} = (q, \chi_1, \chi_2) \simeq (3.0, 0.5, 0.5)$. This case has $\rm{JSD} \simeq 0.022$ for $\delta \Delta t_{\rm{NR}}$ and $\rm{JSD} \simeq 0.017$ for $ \delta d_{\rm{SO}}$. Note that the posterior encompasses zero, indicated by the black lines, which corresponds to the default \seob~fit.}
	\label{fig:pst_example}
\end{figure}

To quantify the similarity between the NR-calibration posteriors residuals, $p(\boldsymbol{\delta \theta} \mid \boldsymbol{\Lambda})$, and our MDN fit, we show in Fig.~\ref{fig:jsd_plot} the Jensen-Shannon divergence (JSD) between 1D marginalized distributions of $\boldsymbol{\delta \theta} = (\delta \Delta t_{\rm{NR}}, \delta d_{\rm{SO}})$, for the 400 spinning NR simulations employed to build the model. The JSD ranges between 0 and 1 bits, with the similarity between the distributions being greater when the JSD is closer to zero. The JSD between two distributions $p(\vartheta)$ and $q(\vartheta)$ is defined as a symmetrized version of the Kullback-Leibler (KL) divergence
\begin{equation}
	\label{eq:JSD}
	\mathrm{JSD} = \frac{D_{\mathrm{KL}}(p | q)  + D_{\mathrm{KL}}(q | p)}{2},
\end{equation}
where the KL divergence is defined as
\begin{equation}
	\label{eq:KL}
	D_{\mathrm{KL}}(p | q) = \int d \vartheta \: p(\vartheta) \log_2 \left(\frac{p(\vartheta)}{q(\vartheta)} \right).
\end{equation}
For both $(\delta \Delta t_{\rm{NR}}, \delta d_{\rm{SO}})$ the median JSD, indicated by the vertical dashed lines, is around $0.02$ bits.

As an illustrative example, we compare in Fig.~\ref{fig:pst_example} the NR-calibration posterior residual $p(\boldsymbol{\delta \theta} \mid \boldsymbol{\Lambda})$ and the corresponding MDN fit, for the NR simulation \texttt{SXS:BBH:2158} with parameters $\boldsymbol{\Lambda} = (q, \chi_1, \chi_2) \simeq (3.0, 0.5, 0.5)$. This is a representative case with $\rm{JSD} \simeq 0.022$ for $\delta \Delta t_{\rm{NR}}$ and $\rm{JSD} \simeq 0.017$ for $ \delta d_{\rm{SO}}$. Note that the posterior encompasses zero, indicated by the black lines, which corresponds to the default \seob~fit. The largest JSD values correspond to broader, more uninformative posteriors, which are harder to fit accurately. These occur particularly near nonspinning or negative-spin configurations. However, this is not a significant issue since uninformative posteriors indicate that the default \seob~model is already accurate.

We now turn to waveform sanity checks, computing the maximum (2,2)-mode mismatch over a range of total masses between 10 and 300 $M_{\odot}$ using the zero-detuned high-power aLIGO PSD~\cite{Barsotti:2018} for different \texttt{SEOBNRv5} variations. We compare the default \texttt{SEOBNRv5} model, a version of \texttt{SEOBNRv5} using the mean of the MDN fit of $p(\boldsymbol{\delta \theta} \mid \boldsymbol{\Lambda})$ for $\boldsymbol{\delta \theta}$, and one in which we take median mismatch over 100 samples for $\boldsymbol{\delta \theta}$ from the MDN model.

\begin{figure}
	\includegraphics[width=0.50\textwidth]{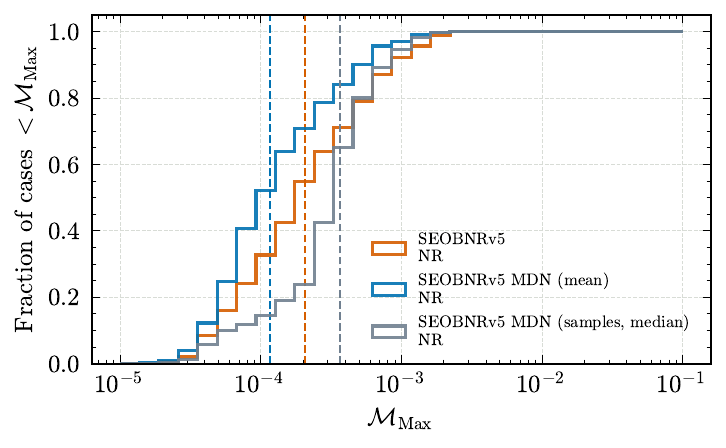}
	\includegraphics[width=0.50\textwidth]{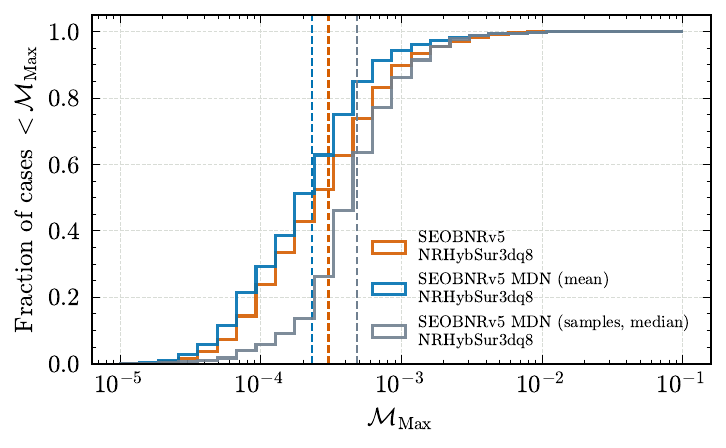}
	\caption{\emph{Top panel:} cumulative distribution of the maximum (2,2)-mode mismatch over 10 and 300 $M_{\odot}$, against the 442 NR simulations used in this work.
		\emph{Bottom panel: } cumulative distribution of the maximum (2,2)-mode mismatch over 10 and 300 $M_{\odot}$, against \texttt{NRHybSur3dq8}, for 5000 random configurations with $q \in [1,8], |\chi_i|\leq 0.9$.
		The vertical dashed lines show the medians. In both panels, we compare the results of the default \texttt{SEOBNRv5} model, a version of \texttt{SEOBNRv5} using the mean of the MDN fit of $p(\boldsymbol{\delta \theta} \mid \boldsymbol{\Lambda})$ for $\boldsymbol{\delta \theta}$, and one in which we take median mismatch over 100 samples for $\boldsymbol{\delta \theta}$ from the MDN model.
	}
	\label{fig:mm_NR_mean}
\end{figure}

In the top panel of Fig.~\ref{fig:mm_NR_mean}, we compare the three \texttt{SEOBNRv5} variations against the 442 NR simulations used in this work, showing the cumulative distribution of the maximum mismatch. We also include 42 nonspinning simulations, not used to build the MDN model, to allow a direct comparison with the results of Ref.~\cite{Pompili:2023tna}.
We observe that using the mean of the MDN fit for $\boldsymbol{\delta \theta}$ results in a more NR-faithful model than the default \texttt{SEOBNRv5}. The median of the distribution, indicated by the vertical dashed lines, decreases by almost a factor of $2$ from $2.1 \times 10^{-4}$ to $1.1 \times 10^{-4}$. Additionally, the fraction of cases with maximum mismatch $\lesssim 1 \times 10^{-3}$ increases from $90\%$ to $96\%$. Using different samples for $\boldsymbol{\delta \theta}$ from the MDN fit also provides comparable accuracy, mostly below the $\sim 1 \times 10^{-3}$ level, consistent with the likelihood function used in the calibration. The median mismatch is $3.6 \times 10^{-4}$, with about $92\%$ of cases having a mismatch $\lesssim 1 \times 10^{-3}$.

In the bottom panel of Fig.~\ref{fig:mm_NR_mean}, we compare the three \texttt{SEOBNRv5} variations to \texttt{NRHybSur3dq8} waveforms, which were not used in the training process. We use 5000 random configurations with $q \in [1,8], |\chi_i|\leq 0.9$ and a dimensionless orbital frequency $M \Omega_0 = 0.015$, allowing some extrapolation outside the surrogate's training region ($|\chi_i|\leq 0.8$) to test the extrapolation of the MDN fit. The mismatch values are comparable to the ones against NR, albeit slightly higher because of the larger number of challenging cases with high $q$ and high spin in this comparison.
The \texttt{SEOBNRv5} version using the mean of the MDN fit for $\boldsymbol{\delta \theta}$ outperforms the default \texttt{SEOBNRv5} model, with the median mismatch decreasing from $3.0 \times 10^{-4}$ to $2.3 \times 10^{-4}$ and fraction of cases with mismatch $\lesssim 1 \times 10^{-3}$ increasing from $87\%$ to $93\%$. This confirms that, regardless of the uncertainty estimate, the MDN fit is a more flexible method compared to the hierarchical least-square fits used in the default \texttt{SEOBNRv5} model. It captures subdominant features present in the NR data without overfitting. Also in this case, taking different samples for $\boldsymbol{\delta \theta}$ from the MDN also provides comparable performance and confirms that the uncertainty estimate leads to well-behaved waveforms overall. The median mismatch in this case is $4.8 \times 10^{-4}$, with about $83\%$ of cases having a mismatch $\lesssim 1 \times 10^{-3}$.

\begin{figure}
	\includegraphics[width=0.50\textwidth]{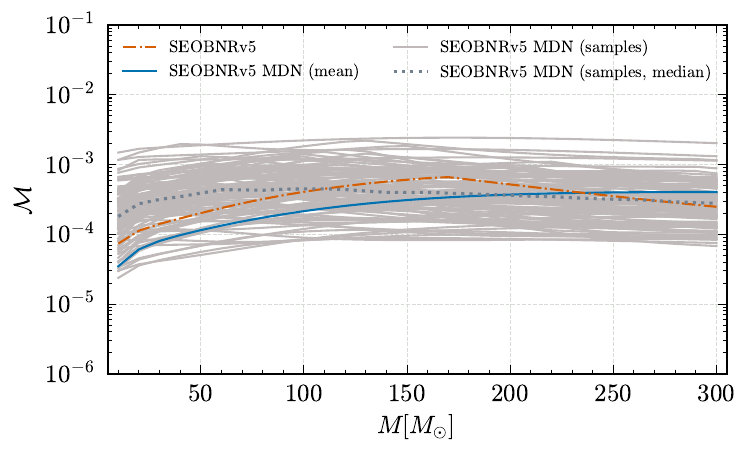}
	\caption{Mismatch distribution as a function of the total mass between \texttt{NRHybSur3dq8} and different variations of \texttt{SEOBNRv5}, for $\boldsymbol{\Lambda} = (q,\chi_1,\chi_2) = (3.0, 0.45, 0.45)$. We show the default \texttt{SEOBNRv5} model and variations which use the mean or different samples of the MDN fit for $\boldsymbol{\delta \theta}$. }
	\label{fig:mm_distr_example}
\end{figure}

Finally, Fig.~\ref{fig:mm_distr_example} shows an example of the mismatch distribution as a function of the total mass for 100 samples from the MDN model. The mismatch is computed against \texttt{NRHybSur3dq8}, with parameters $\boldsymbol{\Lambda} = (q,\chi_1,\chi_2) = (3.0, 0.45, 0.45)$, as the case shown in Fig. 1 of the main text. The darker dotted line indicates the median mismatch across the 100 samples, while the default \texttt{SEOBNRv5} model and the version using the mean of the MDN fit for $\boldsymbol{\delta \theta}$ are represented by the orange and blue lines, respectively.
The mean of the posterior for the calibration parameters generally provides better agreement than most samples, depending on the binary's total mass. However, it does not always correspond to the ``best-fit'' (i.e., maximum likelihood) waveform, as the posterior is not strictly Gaussian. Interpolation of the posteriors across the parameter space also introduces some errors, though this is mitigated by sampling over the $\boldsymbol{\delta \theta}$ parameters during PE.

\section{Impact of higher-modes mismodeling}
\label{app:higher-modes}

\begin{figure}
	\includegraphics[width=0.50\textwidth]{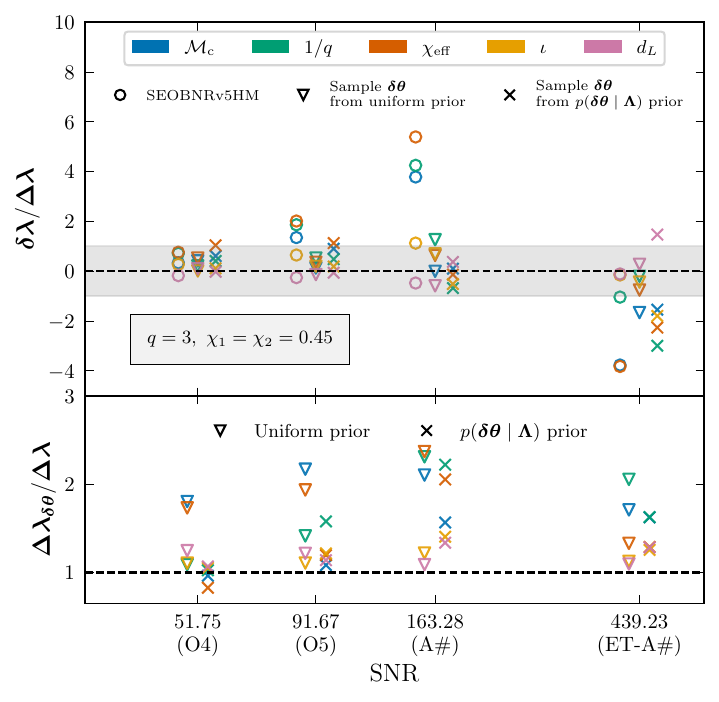}
	\caption{
		Same as the left panel of Fig.~3 in the main text, but for the positive-spin configuration, with $q=3$ and $\chi_i = 0.45$. \emph{Top}: ratio of systematic bias to statistical errors ($\boldsymbol{\delta \lambda} / \boldsymbol{\Delta \lambda}$) as a function of the network SNR. \emph{Bottom}: ratio of statistical errors ($\boldsymbol{\Delta \lambda}$) with and without corrections to the NR-calibration parameters $\boldsymbol{\delta\theta}$, using different priors.
	}
	\label{fig:bias_snr_suppl}
\end{figure}

\begin{figure}
	\includegraphics[width=0.50\textwidth]{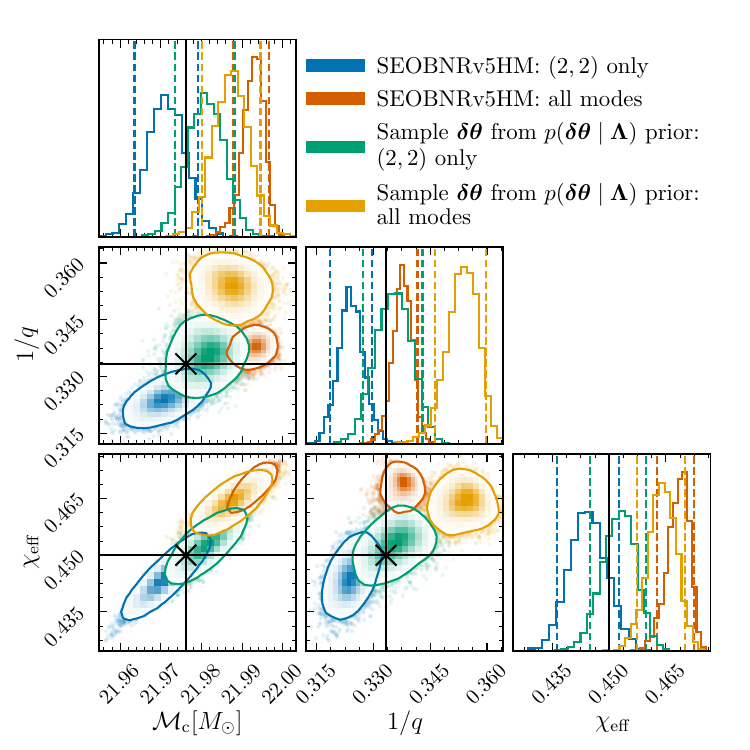}
	\caption{
		Marginalized posterior distributions for the chirp mass, inverse mass ratio, and effective spin in the ET-A\# detector network. Parameter estimation is performed by injecting a \texttt{NRHybSur3dq8} signal and recovering it using \seob, either sampling only the standard source parameters, or including corrections to the NR-calibration parameters $\boldsymbol{\delta\theta}$. For $\boldsymbol{\delta\theta}$, both uniform priors and $p\left(\boldsymbol{\delta \theta} \mid \boldsymbol{\Lambda}\right)$ priors, reflecting NR-calibration uncertainty, are used. We compare results using all available modes in both the signal and template, and using only the dominant $(2,2)$ mode. In the latter case, biases are mitigated when sampling on $\boldsymbol{\delta\theta}$.
	}
	\label{fig:corner_ET_22_only}
\end{figure}

In this appendix, we present PE results for the positive-spin configuration, with $q=3$ and $\chi_i = 0.45$, for upcoming runs, O4, O5, A\#, and next-generation ground-based detector networks (ET-A\#). As for the other runs, PE is performed by injecting a \texttt{NRHybSur3dq8} signal and recovering it using \seob, either sampling only the standard source parameters, or including corrections to the NR-calibration parameters $\boldsymbol{\delta\theta}$, with either uniform or $p\left(\boldsymbol{\delta \theta} \mid \boldsymbol{\Lambda}\right)$ priors, reflecting NR-calibration uncertainty.

Figure~\ref{fig:bias_snr_suppl} is the equivalent of the left panel of Fig.~3 in the main text, but for the positive-spin configuration. It shows the ratio of systematic bias to statistical errors ($\boldsymbol{\delta \lambda} / \boldsymbol{\Delta \lambda}$) as a function of the network SNR in the top panel, and the ratio of statistical errors ($\boldsymbol{\Delta \lambda}$) with and without corrections to the NR-calibration parameters $\boldsymbol{\delta\theta}$, in the bottom panel.
For the O5 and A\# detector networks, where the injected signal has SNR $\sim 100$, accounting for NR-calibration uncertainties by sampling over the $\boldsymbol{\delta\theta}$ parameters substantially mitigates the biases recovered by the \seob~model. This is not true in the ET-A\# network, in which the injected signal has SNR $= 439.2$, although biases are still slightly reduced, especially when looking at 2D marginalized posterior distributions. The presence of a bias suggests that the $\boldsymbol{\delta\theta}$ parameters do not provide sufficient flexibility to match \texttt{NRHybSur3dq8} with the required precision, for the same values of the physical parameters $\boldsymbol{\lambda}$.
This could be due to the ineffectiveness of the parameters being varied, missing physics in the \seob~model (e.g., some higher-order modes), or potential inaccuracies in the \texttt{NRHybSur3dq8} model (either due to NR resolution or to the hybridization procedure), which may become significant at such high SNRs~\cite{Purrer:2019jcp}. This is expected, as the chosen $\boldsymbol{\delta\theta}$ may not fully account for all sources of error, and mismodeling in other parts of the waveform construction could become relevant at sufficiently high SNR.

One possible explanation for the residual biases is the absence or mismodeling of higher-order modes.
In the left panel of Fig.~3 in the main text, we note that the configuration in which higher modes are most excited ($q=6$ and $\chi_i=0.75$, with an SNR in the $(3,3)$ mode of approximately $28$) results in moderate biases even for the model incorporating uncertainties -- although these biases are considerably reduced compared to the default model. For the configuration with $q=3$ and $\chi_i=0.45$ in the ET-A\# network, which has a higher overall SNR, the SNR in the $(3,3)$ mode is about $43$, potentially contributing to a significant impact from higher-mode mismodeling.

To quantify the impact of neglecting certain higher-order modes, we first perform a test where we include the same set of modes in both the injection and recovery waveform models. Specifically, \texttt{NRHybSur3dq8} includes the $(\ell,m) = (2,0), (3,1), (3,0), (4,2), (5,5)$ modes, which are not incorporated by default in \seob. We find that this change reduces the biases only slightly, and does not qualitatively affect the results, indicating that the lack of higher-order modes is not the primary source of error.

In \seob, higher-order modes are calibrated to NR in the merger-ringdown phase, but they are less accurate than the dominant $(2,2)$ mode~\cite{Pompili:2023tna}. This is due to both the lower NR quality for these modes and the increased difficulty of fitting subdominant modes, which often exhibit complex morphologies. The $\boldsymbol{\delta\theta}$ parameters we focus on do not account for these inaccuracies. To assess the impact of higher-mode mismodeling, we repeat the analysis using only the dominant $(2,2)$ mode in both the signal and template. We show in Fig.~\ref{fig:corner_ET_22_only} the marginalized posterior distributions for the chirp mass, inverse mass ratio, and effective spin. In this case, we observe that biases are still present in \seob, but are mitigated when sampling on $\boldsymbol{\delta\theta}$. This indicates that mismodeling of higher-order modes is a significant source of error.
The NR-calibration parameters $\boldsymbol{\theta}$ we consider only shape the waveform model in specific directions; it appears that the waveform space that these parameters allow to explore that does not allow us to approach the NR waveform closely enough without also modifying the standard binary parameters $\boldsymbol{\lambda}$. 
To address this, additional $\boldsymbol{\delta\theta}$ parameters should be considered to account for mismodeling specific to higher-order modes. This could be achieved by introducing parametrized deviations in the amplitude and frequency at the merger, similar to what is done in tests of GR~\cite{Maggio:2022hre}, after determining appropriate priors across the parameter space.

\section{Impact of NR length on calibration}

\begin{figure}[t]
	\includegraphics[width=0.50\textwidth]{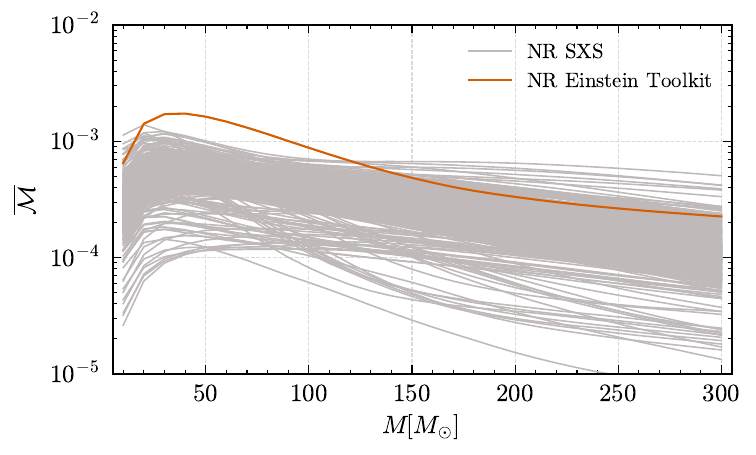}
	\includegraphics[width=0.50\textwidth]{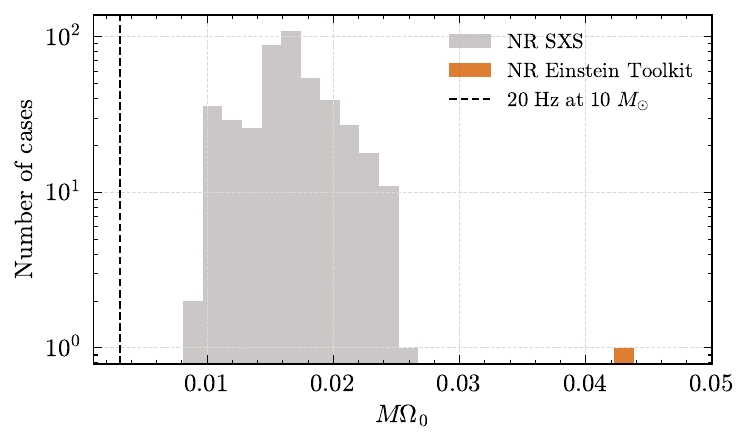}
	\caption{
		\emph{Top panel: } median mismatch between \texttt{SEOBNRv5} using for the calibration parameters $\boldsymbol{\theta}$ the calibration posterior median, and \texttt{SEOBNRv5} with 100 posterior samples, starting from a frequency of $20~\rm{Hz}$ for $10 M_{\odot}$. Each curve corresponds to the calibration posterior obtained from one of the 400 spinning NR simulations used in this work.
		\emph{Bottom panel: } dimensionless initial frequency of the NR waveforms used in this work. We highlight in orange the $q=8$ \texttt{Einstein Toolkit} simulation, which covers $7.4$ orbits and is the shortest we employ. The black dashed line indicates the frequency at which we start the mismatch computation in this section.}
	\label{fig:NR_low_freq}
\end{figure}

In this section, we investigate whether the NR calibration of \texttt{SEOBNRv5} (and its associated uncertainty), which is only informative of the accuracy of the model at frequencies larger than the initial NR starting frequency for each simulation, is sufficient to constrain the low-frequency portion of the waveform, not covered by NR simulations. Since direct comparison with NR is not possible in this regime, we perform internal consistency tests. Specifically, we assess if different values of the EOB calibration parameters $\boldsymbol{\theta}$ that faithfully reproduce an NR waveform with parameters $\boldsymbol{\Lambda}$ at high frequencies [e.g., different calibration posterior samples from $p\left(\boldsymbol{\theta} \mid \boldsymbol{\Lambda}\right)$] can result in significant variations in the low-frequency behavior.
To do so, we compute the unfaithfulness between \texttt{SEOBNRv5} taking a reference value for the calibration parameters $\boldsymbol{\theta}$ (e.g., the posterior median) and \texttt{SEOBNRv5} using different posterior samples. We compute the average unfaithfulness $\bar{\mathcal{M}}$ over 100 samples for each NR case, starting from a frequency of $20~\rm{Hz}$ for $10 M_{\odot}$ (corresponding to a dimensionless frequency $M \Omega_0 \simeq 0.003$), which is much lower than the typical NR starting frequency, as shown in the bottom panel of Fig.~\ref{fig:NR_low_freq}. This was previously investigated for the \texttt{SEOBNRv4} model in Sec. 6 of Ref.~\cite{Bohe:2016gbl} (see, e.g., Fig. 7), starting however at a larger frequency of $25~\rm{Hz}$ for $20 M_{\odot}$.

The top panel Fig.~\ref{fig:NR_low_freq} shows the median mismatch as a function of the total mass. The unfaithfulness generally remains below the $10^{-3}$ level, which is comparable to the mismatch against NR using different posterior samples in the high-frequency region.
This means that different values of the calibration parameters do not significantly affect the low-frequency behavior, even for very short NR waveforms (e.g., the $q=8$ \texttt{Einstein Toolkit} waveform covering only $7.4$ orbits). This indicates that the low-frequency behavior of the model is robust, with most uncertainty arising from the late-inspiral and merger portions of the signal covered by NR.
In contrast, for \texttt{SEOBNRv4}, the mismatch could exceed $10\%$ for the $q=8$ \texttt{Einstein Toolkit} simulation. The better behavior of \texttt{SEOBNRv5} is likely due to having fewer calibration parameters compared to \texttt{SEOBNRv4}, which are only needed to correct the final portion of the waveform, thanks to a more accurate analytic baseline model.

\section{Bayes factors and posterior distributions}
\label{app:BF}

Besides posterior distributions, nested sampling~\cite{Skilling:2006gxv} provides an estimate of the Bayesian evidence $Z \equiv \int \mathrm{d} \Theta \mathcal{L}(d \mid \Theta) \pi(\Theta)$, where $\mathcal{L}(d|\Theta)$ is the likelihood of the data $d$ given the parameters $\Theta$, $\pi(\Theta)$ is the prior on $\Theta$.
To determine whether a model $A$ is preferred over a model $B$, one can look at the Bayes factor, defined as the ratio of the evidence for the two different models $\mathcal{B}^A_B=Z_A/Z_B$~\cite{Thrane:2018qnx}.

The Bayes factor naturally accounts for model complexity: models with more parameters tend to fit the data better and achieve a higher likelihood, but they are penalized due to the larger prior volume. In our study, this allows us to quantify both the improvement due to the additional $\boldsymbol{\delta\theta}$ parameters and the impact of choosing uniform or $p\left(\boldsymbol{\delta \theta} \mid \boldsymbol{\Lambda}\right)$ priors. Here, we take $\Theta = \boldsymbol{\lambda}$ for the default \texttt{SEOBNRv5HM} model and $\Theta = \boldsymbol{\lambda} \cup \boldsymbol{\delta\theta}$ for the models that include corrections to the NR-calibration parameters.

In this Appendix, we report natural log Bayes factors between models with corrections to the NR-calibration parameters $\boldsymbol{\delta \theta}$ and the default \texttt{SEOBNRv5HM} model, with positive values indicating a preference for the models including corrections. For all cases, the estimated uncertainties on $\ln{\mathcal{B}}$ is approximately $\pm 0.3$. Specifically, we report the following results:
\begin{table}[t]
	\centering
	\begin{tabular}{c|c|c|c|c}
		\hline
		\hline
		$(q,~\chi_i)$ & Network & SNR   & \multicolumn{2}{c}{$\ln{\mathcal{B}}$}                                                       \\
		\cmidrule(lr){3-4}
					&         &       & $p\left(\boldsymbol{\delta \theta} \mid \boldsymbol{\Lambda}\right)$ priors & Uniform priors \\
		\hline
		($3,~0.45$)   & O4      & 51.8  & -0.2                                                                        & -1.8           \\
		($3,~0.45$)   & O5      & 91.7  & 0.4                                                                         & -1.9           \\
		($3,~0.45$)   & A\#     & 163.3 & 2.1                                                                         & -0.3           \\
		($3,~0.45$)   & ET-A\#  & 439.2 & 31.7                                                                        & 27.1           \\
		\hline
		($3,~-0.45$)  & O4      & 39.3  & 0.1                                                                         & -0.3           \\
		($3,~-0.45$)  & O5      & 70.0  & 0.4                                                                         & -0.5           \\
		($3,~-0.45$)  & A\#     & 125.6 & 0.6                                                                         & 0.4            \\
		($3,~-0.45$)  & ET-A\#  & 366.7 & 6.8                                                                         & 7.3            \\
		\hline
		\hline
	\end{tabular}
	\caption{Natural log Bayes factors between models with corrections to the NR-calibration parameters $\boldsymbol{\delta \theta}$ and the default \texttt{SEOBNRv5HM} model, for \texttt{NRHybSur3dq8} injections in different detector network. These analyses incorporate all higher modes present in the models.}
	\label{tab:bayes_1}
\end{table}
\begin{table}[t]
	\centering
	\begin{tabular}{c|c|c|c}
		\hline
		\hline
		$(q,~\chi_i)$  & Network & SNR   & $\ln{\mathcal{B}}~(p\left(\boldsymbol{\delta \theta} \mid \boldsymbol{\Lambda}\right)~\text{priors})$ \\
		\hline
		($1.5,~-0.75$) & A\#     & 130.9 & 1.0                                                                                                   \\
		($1.5,~-0.45$) & A\#     & 140.0 & -0.2                                                                                                  \\
		($1.5,~0.45$)  & A\#     & 173.6 & -0.5                                                                                                  \\
		($1.5,~0.75$)  & A\#     & 187.8 & 0.5                                                                                                   \\
		($3,~-0.75$)   & A\#     & 116.3 & -0.4                                                                                                  \\
		($3,~-0.45$)   & A\#     & 125.6 & 0.6                                                                                                   \\
		($3,~0.45$)    & A\#     & 163.3 & 2.1                                                                                                   \\
		($3,~0.75$)    & A\#     & 180.7 & -0.1                                                                                                  \\
		($6,~-0.75$)   & A\#     & 88.0  & 0.3                                                                                                   \\
		($6,~-0.45$)   & A\#     & 98.3  & 1.9                                                                                                   \\
		($6,~0.45$)    & A\#     & 144.9 & -1.0                                                                                                  \\
		($6,~0.75$)    & A\#     & 167.2 & 4.0                                                                                                   \\
		\hline
		\hline
	\end{tabular}
	\caption{Natural log Bayes factors between models with corrections to the NR-calibration parameters $\boldsymbol{\delta \theta}$ and the default \texttt{SEOBNRv5HM} model, for \texttt{NRHybSur3dq8} injections in the A\# detector network. These analyses incorporate all higher modes present in the models.}
	\label{tab:bayes_2}
\end{table}
\begin{table}[t]
	\centering
	\begin{tabular}{c|c|c|c|c}
		\hline
		\hline
		$(q,~\chi_i)$ & Network & SNR   & Modes     & $\ln{\mathcal{B}}~(p\left(\boldsymbol{\delta \theta} \mid \boldsymbol{\Lambda}\right)~\text{priors})$ \\
		\hline
		($3,~0.45$)   & ET-A\#  & 439.2 & All modes & 31.7                                                                                                  \\
		($3,~0.45$)   & ET-A\#  & 428.1 & $(2,2)$   & 7.2                                                                                                   \\
		\hline
		\hline
	\end{tabular}
	\caption{Natural log Bayes factors between models with corrections to the NR-calibration parameters $\boldsymbol{\delta \theta}$ and the default \texttt{SEOBNRv5HM} model, for \texttt{NRHybSur3dq8} injections in the ET-A\# detector network with and without higher modes.}
	\label{tab:bayes_3}
\end{table}

\begin{itemize}
	\item Table~\ref{tab:bayes_1}: natural log Bayes factors between the models with $\boldsymbol{\delta \theta}$ corrections and the default \texttt{SEOBNRv5HM} model, for \texttt{NRHybSur3dq8} injections in different detector networks (left panel of Fig.~3 in the main text and Fig.~\ref{fig:bias_snr_suppl}). Results are shown for both \(p\left(\boldsymbol{\delta\theta} \mid \boldsymbol{\Lambda}\right)\) and uniform priors. These analyses incorporate all higher modes present in the models.

	\item Table~\ref{tab:bayes_2}: natural log Bayes factors for \texttt{NRHybSur3dq8} injections in the A\# detector network with various parameters, corresponding to the right panel of Fig.~3 in the main text. In this case, only results using \(p\left(\boldsymbol{\delta\theta} \mid \boldsymbol{\Lambda}\right)\) priors are provided. These analyses incorporate all higher modes present in the models.

	\item Table~\ref{tab:bayes_3}: natural log Bayes factors for \texttt{NRHybSur3dq8} injections in the ET-A\# detector network with and without higher modes, corresponding to Fig~\ref{fig:corner_ET_22_only}.
\end{itemize}

From Table~\ref{tab:bayes_1}, we see that Bayes factors tend to increase with more sensitive detector networks, particularly for the case with positive spins, indicating that the inclusion of NR-calibration corrections is more strongly favored. This trend is observed even in the ET-A\# detector network, despite the recovered parameters being biased. This implies that the $\boldsymbol{\delta\theta}$ parameters allow for a better match to \texttt{NRHybSur3dq8}, although not necessarily for the same values of the physical parameters $\boldsymbol{\lambda}$.
The choice of prior also has a moderate impact on the Bayes factors: models employing \(p\left(\boldsymbol{\delta\theta} \mid \boldsymbol{\Lambda}\right)\) priors yield mostly positive Bayes factors -- improving the fit to the data with little increase in prior volume -- whereas more uninformative, uniform priors can lead to slightly negative Bayes factors, especially for the O4 and O5 detector networks.
Focusing on the A\# detector network in Table~\ref{tab:bayes_2}, where we explore a broader range of binary parameters, we observe that some configurations show a moderate preference toward the model with for \(p\left(\boldsymbol{\delta\theta} \mid \boldsymbol{\Lambda}\right)\) priors, while others indicate little or no improvement over the default \texttt{SEOBNRv5HM} model. This is roughly consistent with Fig.~3, which shows that in some cases the default model is not significantly biased.

\begin{figure*}
	\includegraphics[width=\columnwidth]{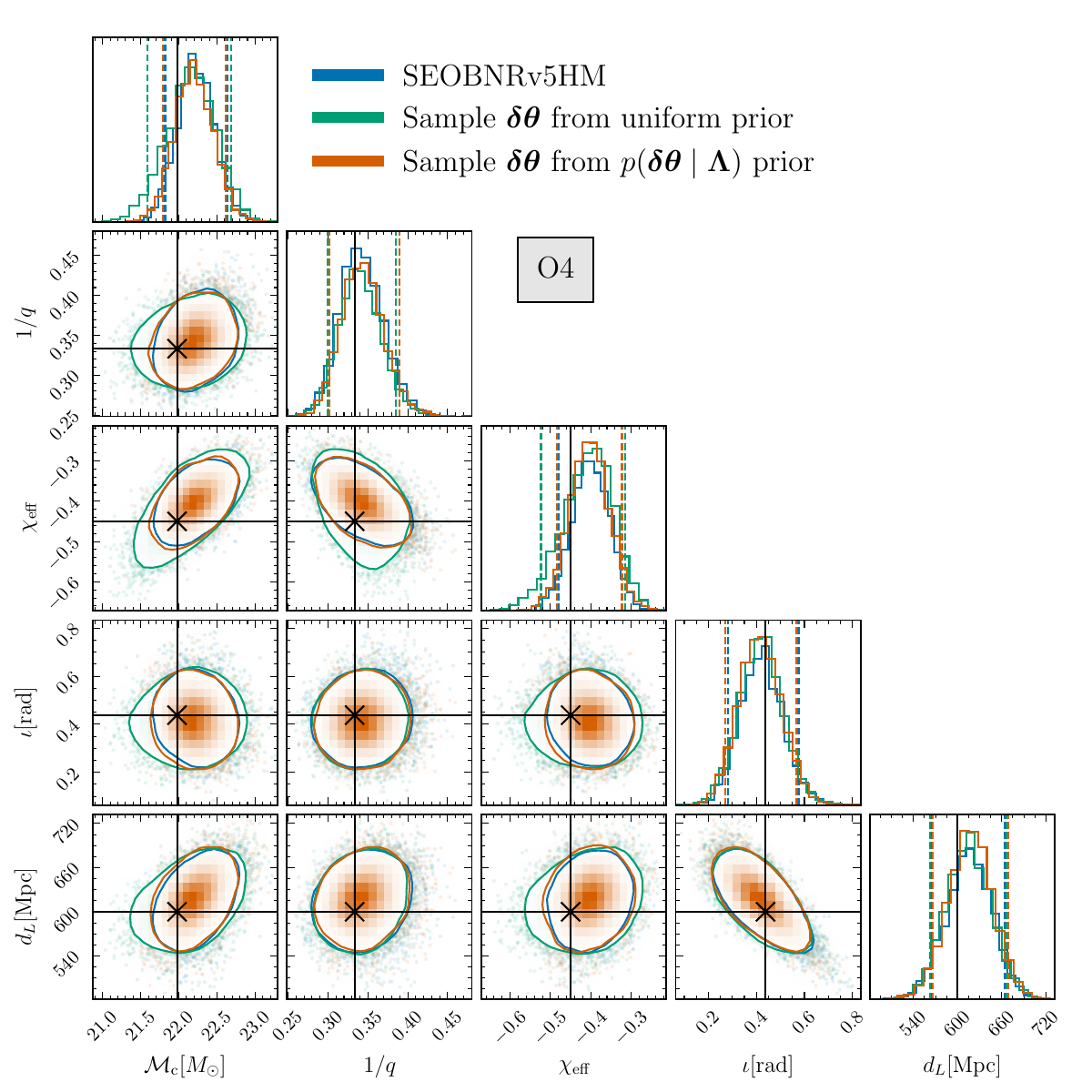}
	\includegraphics[width=\columnwidth]{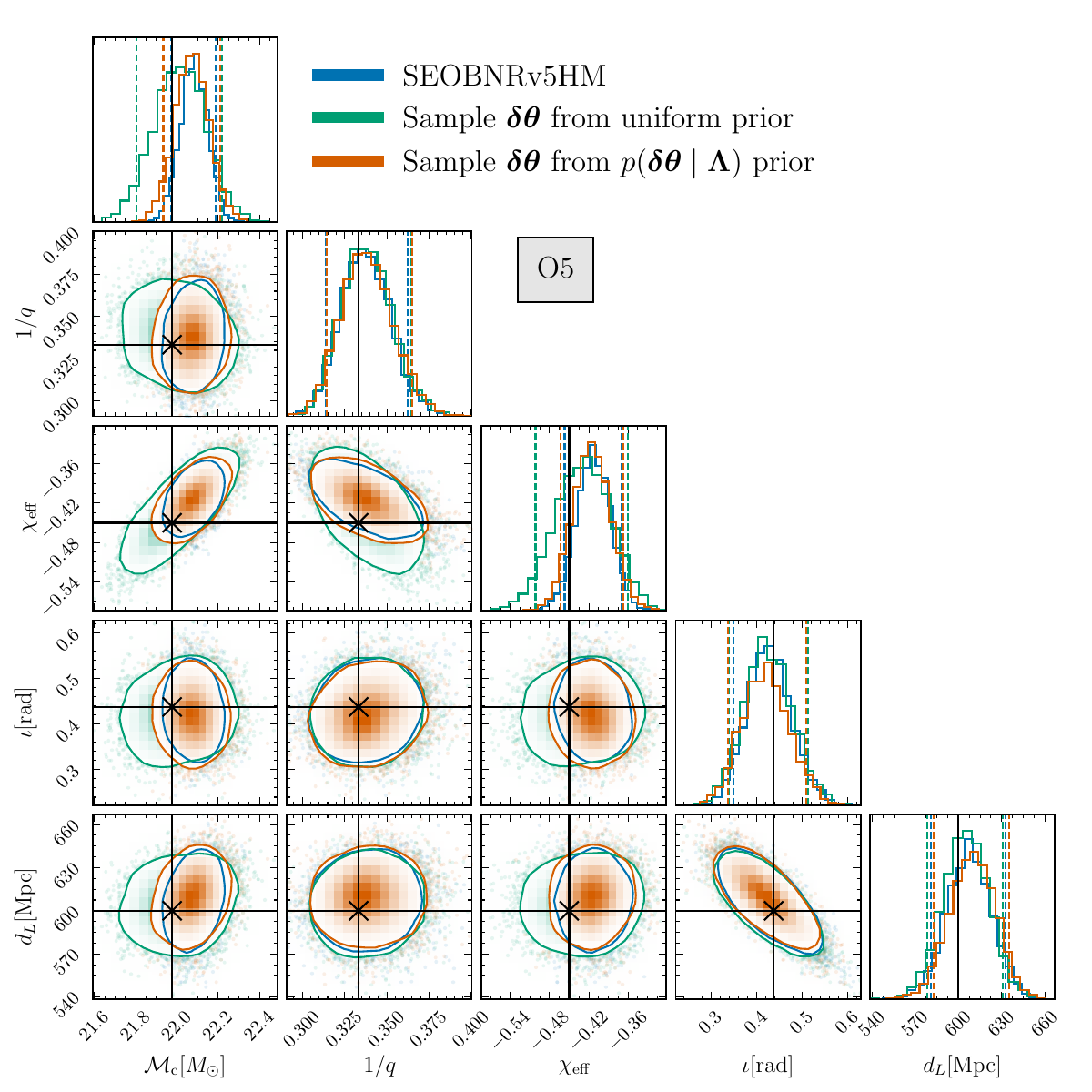}
	\includegraphics[width=\columnwidth]{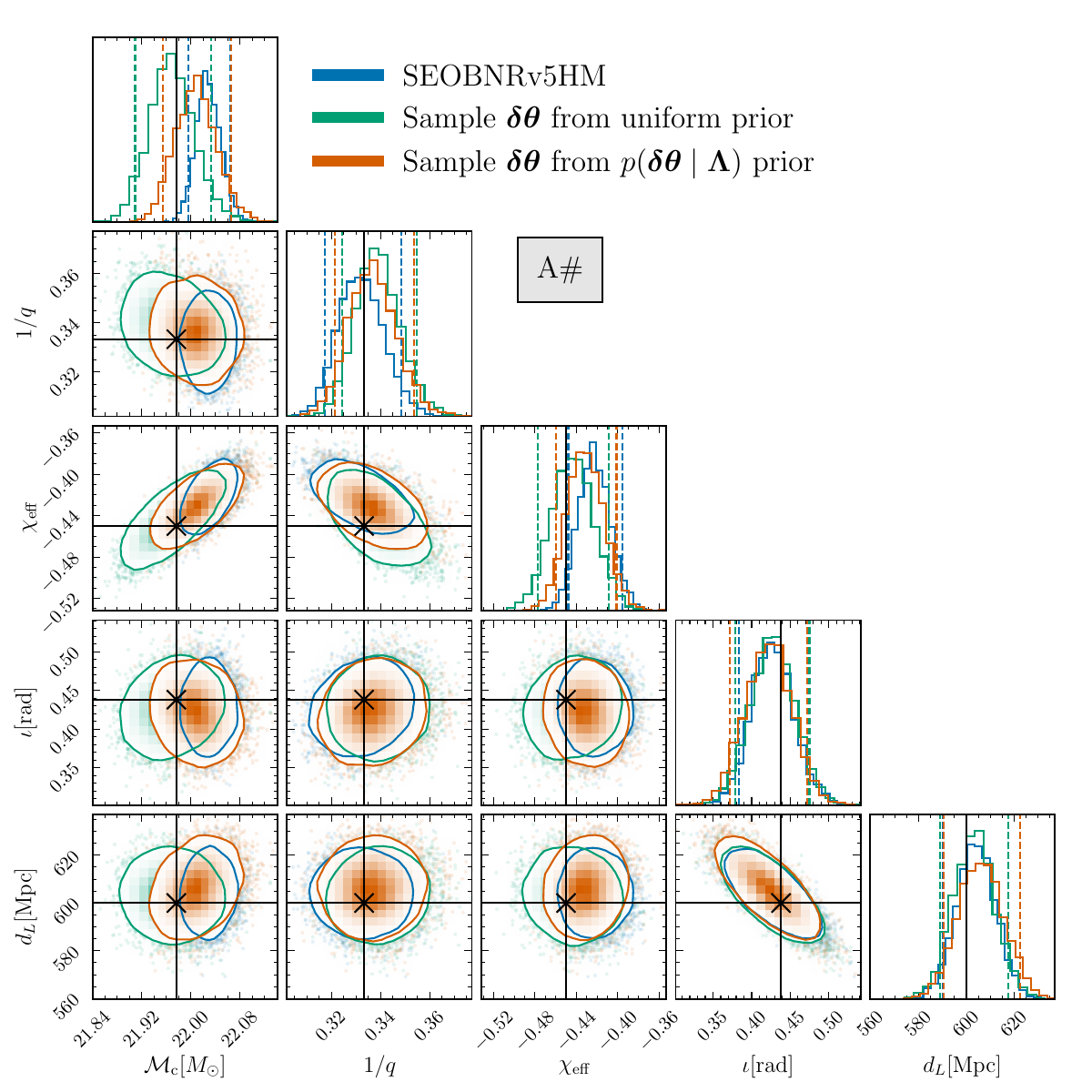}
	\includegraphics[width=\columnwidth]{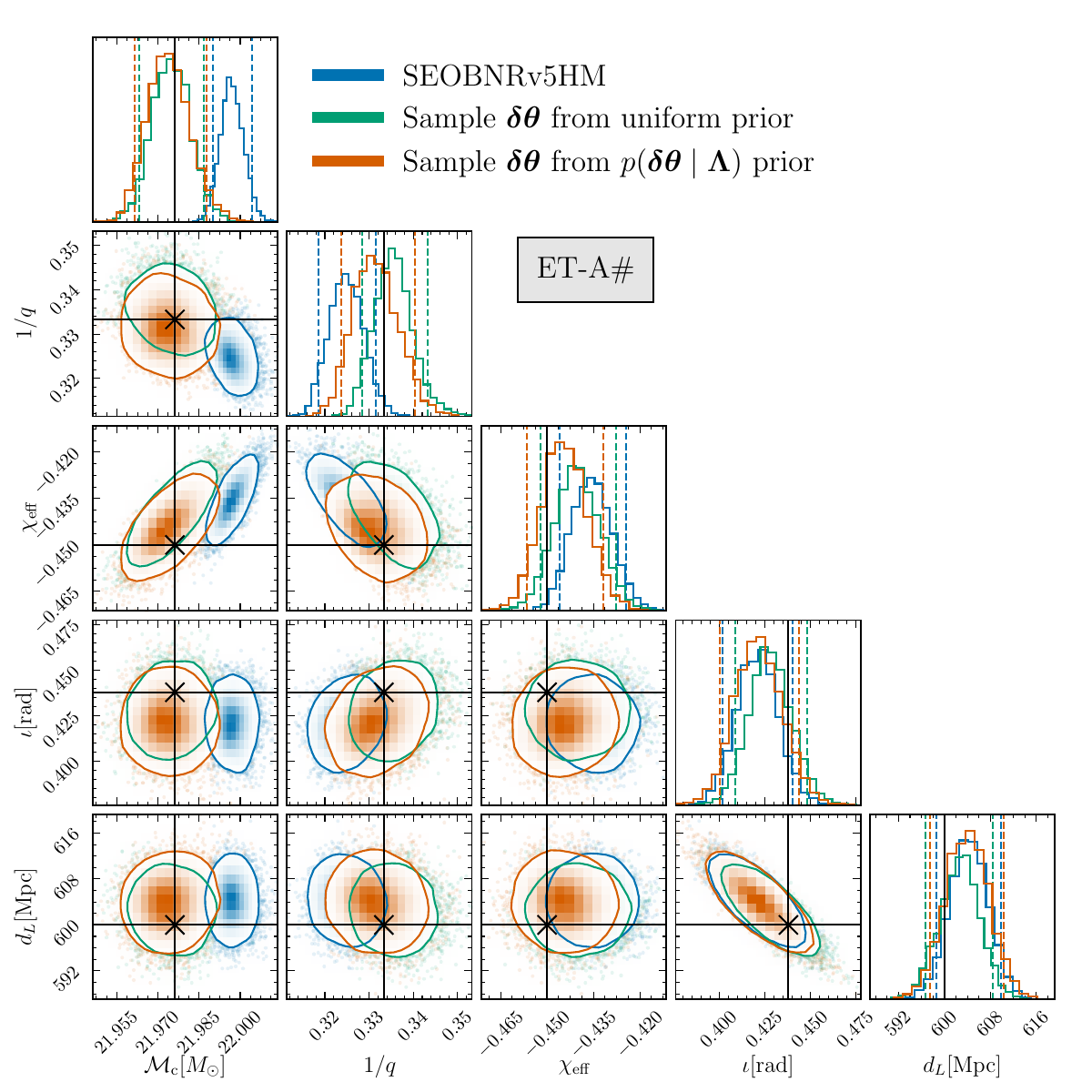}
	\caption{Marginalized posterior distributions for the chirp mass, inverse mass ratio, effective spin, inclination, and luminosity distance, in the different detector networks. Parameter estimation is performed by injecting a \texttt{NRHybSur3dq8} signal, with unequal masses ($q=3$) and negative spins ($\chi_1 = \chi_2 = - 0.45$), and recovering it using three versions of \seob: one sampling only the standard parameters, and the others including corrections to the NR-calibration parameters $\boldsymbol{\delta\theta}$. For $\boldsymbol{\delta\theta}$, both uniform priors and $p\left(\boldsymbol{\delta \theta} \mid \boldsymbol{\Lambda}\right)$ priors, reflecting NR-calibration uncertainty, are used.  }
	\label{fig:corner_1}
\end{figure*}

\begin{figure*}
	\includegraphics[width=\columnwidth]{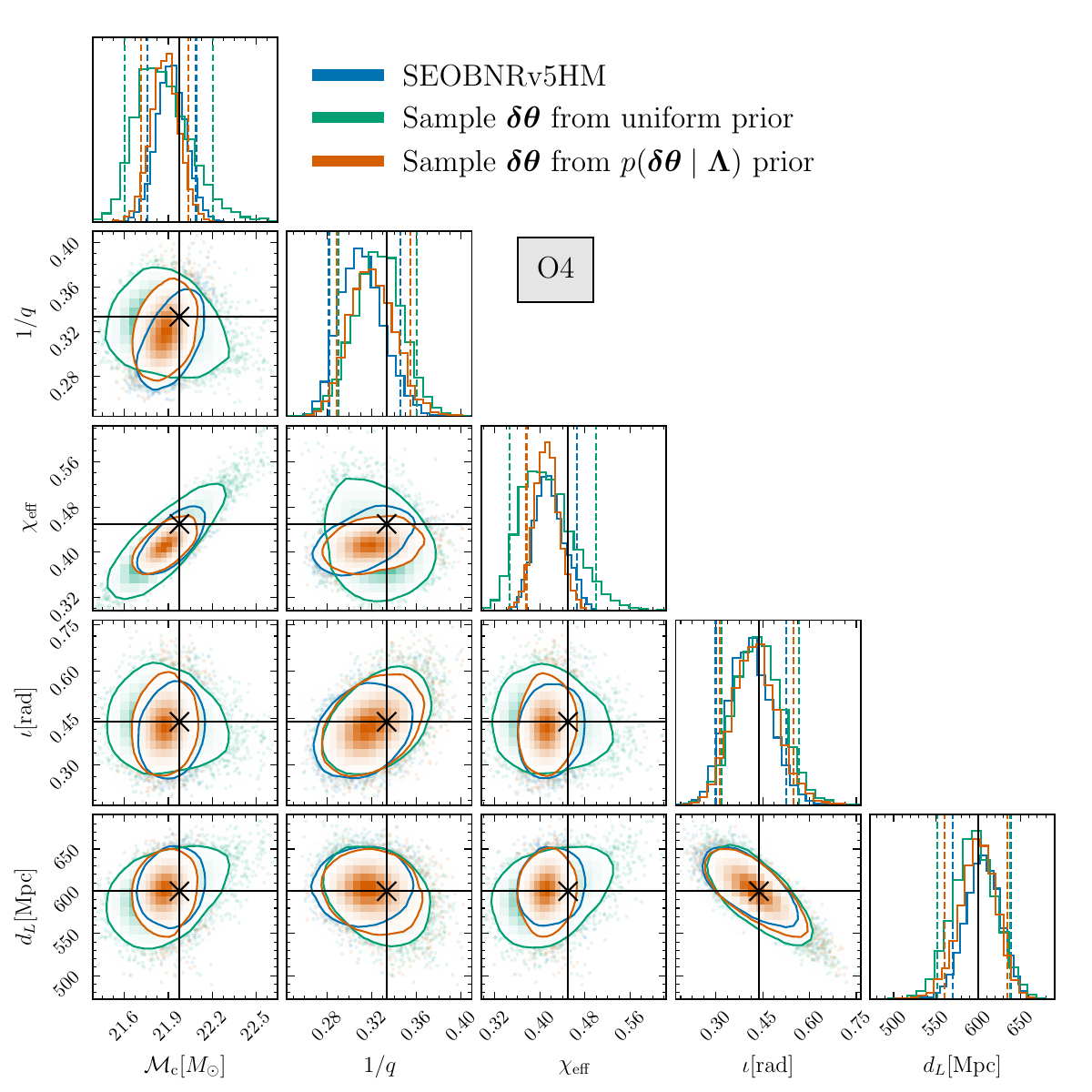}
	\includegraphics[width=\columnwidth]{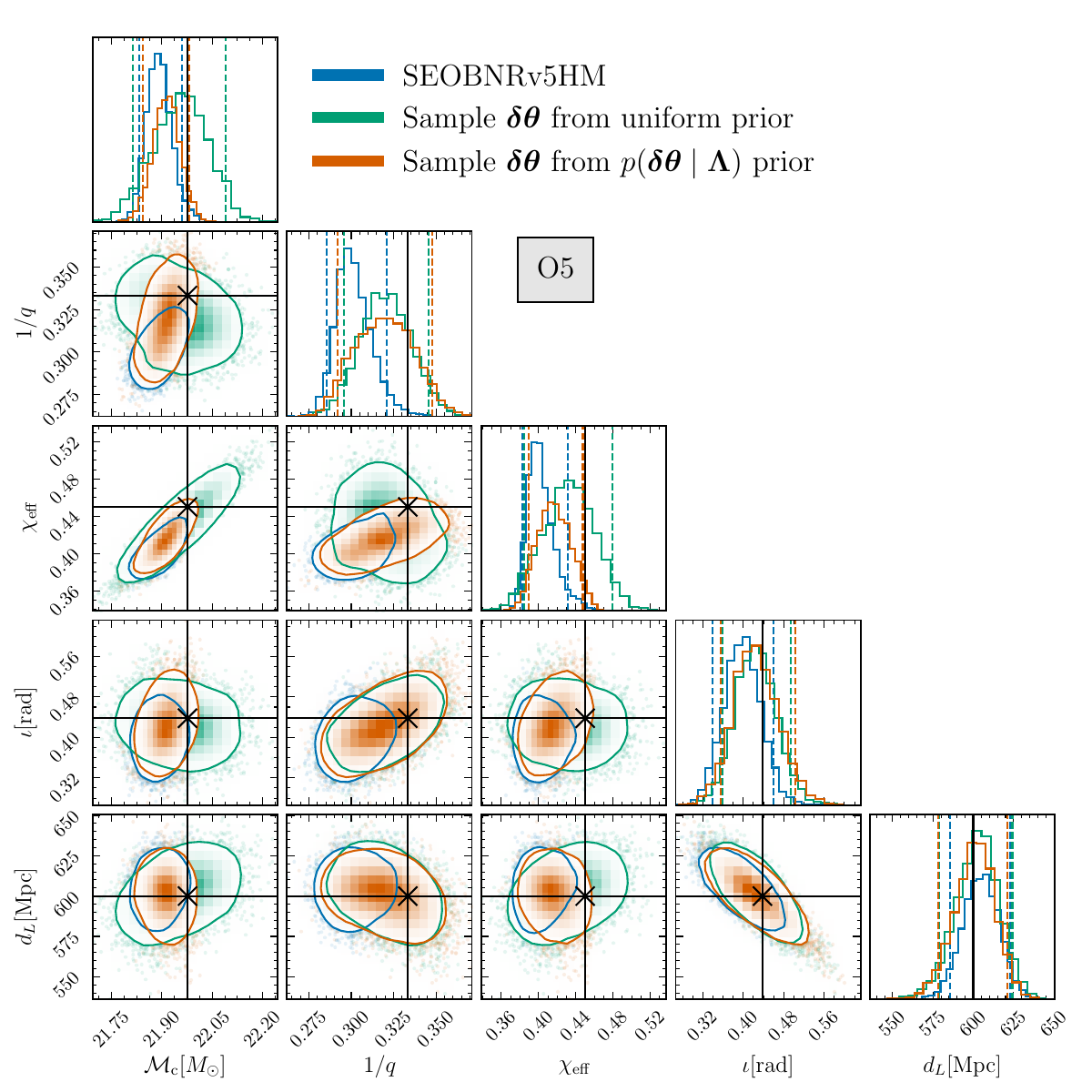}
	\includegraphics[width=\columnwidth]{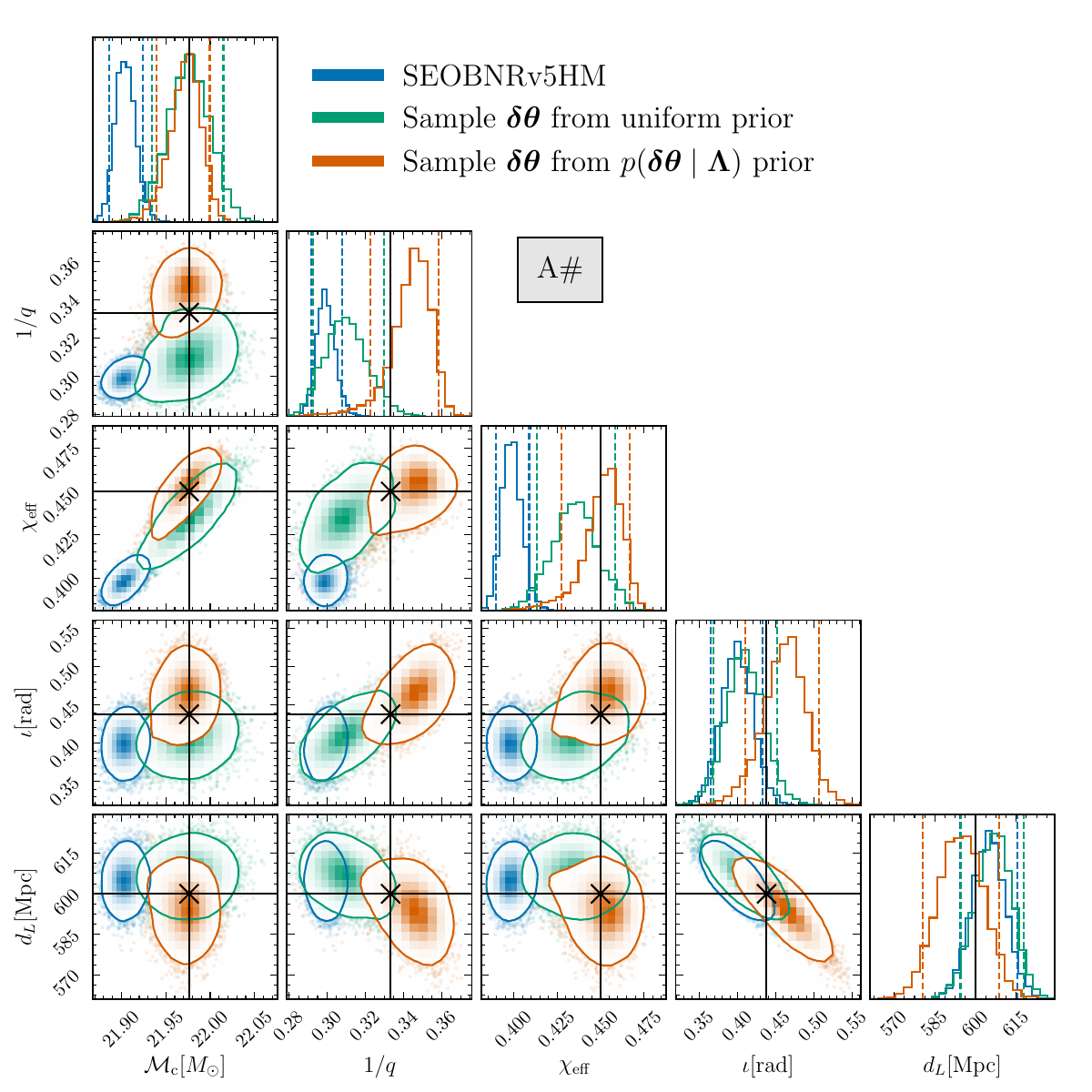}
	\includegraphics[width=\columnwidth]{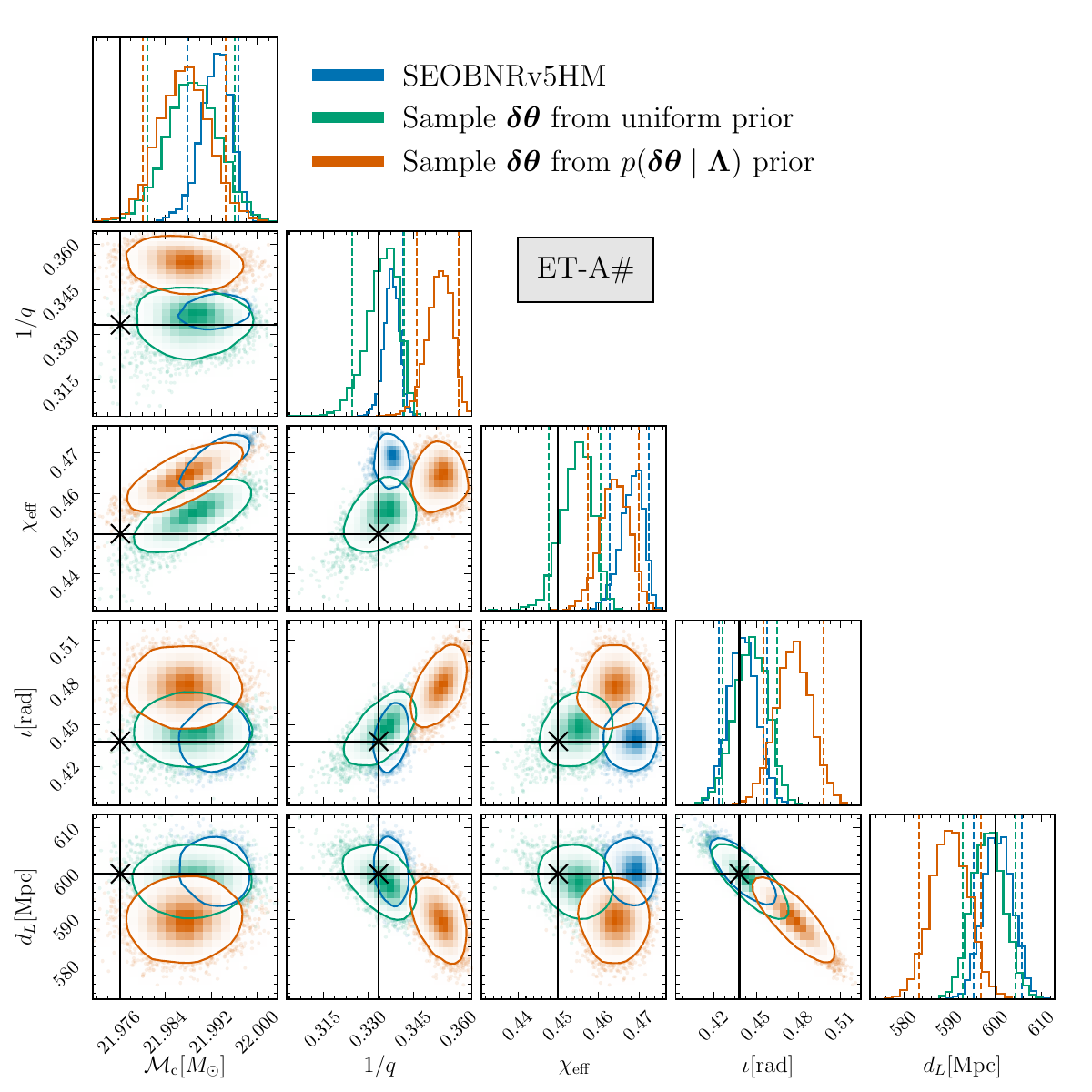}
	\caption{Marginalized posterior distributions for the chirp mass, inverse mass ratio, effective spin, inclination, and luminosity distance, in the different detector networks. Parameter estimation is performed by injecting a \texttt{NRHybSur3dq8} signal, with unequal masses ($q=3$) and positive spins ($\chi_1 = \chi_2 = 0.45$), and recovering it using three versions of \seob: one sampling only the standard parameters, and the others including corrections to the NR-calibration parameters $\boldsymbol{\delta\theta}$. For $\boldsymbol{\delta\theta}$, both uniform priors and $p\left(\boldsymbol{\delta \theta} \mid \boldsymbol{\Lambda}\right)$ priors, reflecting NR-calibration uncertainty, are used. }
	\label{fig:corner_0}
\end{figure*}

For completeness, we also show in Figs.~\ref{fig:corner_1} and~\ref{fig:corner_0} the marginalized posterior distributions for the parameter recoveries of the configurations with $q=3$ and $\chi_i = \pm 0.45$ across the different detector networks.
This provides a complementary picture to Fig.~3 in the main text and Fig.~\ref{fig:bias_snr_suppl} here, which focus only on biases in the 1D marginalized posteriors, by illustrating correlations between parameters and the appearance of biases in the 2D marginalized posteriors. This also serves to highlight the significant improvement in measurement precision with higher SNR signals seen by upcoming GW detectors.

\end{document}